# Salt-induced gelation

# of nonionic sucrose ester dispersions

D. Cholakova[1], N. Pagureva[1], M. Hristova[1,2], S. Tcholakova[1]*

*[1]Department of Chemical and Pharmaceutical Engineering*

*Faculty of Chemistry and Pharmacy*

*Sofia University, 1164 Sofia, Bulgaria*

*[2]Centre of Competence "Sustainable Utilization of Bio-resources and Waste of Medicinal and Aromatic Plants for Innovative Bioactive Products" (BIORESOURCES BG), Sofia, Bulgaria*

*__Corresponding author__:

Prof. Slavka Tcholakova

Department of Chemical and Pharmaceutical Engineering

Faculty of Chemistry and Pharmacy, Sofia University

1 James Bourchier Ave., 1164 Sofia

Bulgaria

Phone: (+359-2) 816 1698

E-mail: sc@lcpe.uni-sofia.bg

**Abstract**

**Hypothesis:**

The dispersions of nonionic sucrose ester surfactants in water exhibit a highly negative zeta-potential, though its origin remains controversial. The addition of electrolytes to these dispersions may influence their zeta-potential, thus potentially affecting their physicochemical properties.

**Experiments:**

The electrolyte- and pH- driven gelation of aqueous dispersions of commercial sucrose stearate (S970) containing ca. 1:1 monoesters and diesters was studied using optical microscopy, rheological and zeta-potential measurements, and small-angle X-ray scattering techniques.

**Findings:**

At low electrolyte concentrations and pH $\gtrsim$ 5, 0.5-5 wt. % S970 dispersions exhibited low viscosities and behaved as freely flowing liquids. The addition of electrolytes of low concentrations, e.g. 9 mM NaCl or 1.5 mM MgCl$_2$, induced the formation of a non-flowing gels. This sol-gel transition occurred due to the partial screening of the diesters particles charge, allowing the formation of an attractive gel network, spanning across the dispersion volume. Complete charge screening, however, led to a gel-sol transition and phase separation. Gel formation was observed also by pH variation without electrolyte addition, whereas the addition of free fatty acids had negligible impact on dispersion properties. These findings support the hypothesis that the negative charge in sucrose ester dispersions arises from hydroxyl anions adsorption on particles surfaces. Gels were formed using just 1.3 wt. % surfactant, and the critical electrolyte concentration for gelation was found to scale approximately with the square of the cation charge, in agreement with the low surface charge density theory. The biodegradable sucrose esters gels offer a sustainable alternative for structuring personal and home care products, replacing the wormlike micelles of synthetic surfactants typically used at much higher surfactant and salt concentrations.

## 1. Introduction

The fine tuning of the rheological properties of surfactant solutions is an important aspect of product formulation in personal and home care products, cosmetics, pharmaceutics, food science, oil recovery and many other areas [1-6]. Formation of disperse systems with increased viscosities or even gel-like properties is desired in many cases. Gelation can be achieved through diverse interactions between the dispersed molecules or particles and the solvent, which lead to the disperse medium localization in space and emergence of macroscopic rigidity of the system [7]. Rheological properties are often controlled by the addition of various polymeric molecules which cross-links by either physical (hydrogen bonding, hydrophobic interactions, inter-chain entanglement interactions, local crystallite formation) or chemical (covalent bonds) interactions [7].

In absence of actual polymers, an alternative route for viscosity modification in surfactant containing systems is to use their self-assembly properties. For example, it is well established that the formation of wormlike micelles leads to the appearance of viscoelastic properties in the surfactant solutions [8-10]. Owing to their similarity with the high molecular weight macromolecules and their dynamic nature, wormlike micelles are even referred to as "living polymers" in the literature [11,12]. Wormlike micelles are typically formed in ionic and zwitterionic surfactant systems when the electrostatic repulsion between the charged hydrophilic headgroups is appropriately screened. One typical example are the entangled wormlike micelles formed when the anionic sodium lauryl ether sulfate (SLES) surfactant is mixed with the zwitterionic cocoamidopropylbetaine (CAPB) in presence of salts or other small molecules acting as co-surfactants [10]. In presence of 10 wt. % SLES + CAPB (2:1 w/w) and 1 wt. % octanol as co-surfactant, zero-shear viscosity exceeding $10^3$ Pa.s has been reported [10]. Due to its excellent foaming and cleansing properties, this combination of surfactants is widely used in many cosmetic products. However, note that relatively high surfactant concentrations (typically > 8-10 wt. %) are needed to obtain non-flowing gel-like samples structured solely with surfactant molecules.

An alternative approach for formation of highly viscous and non-flowing samples is the formation of space-spanning network within the disperse system composed of interacting particles. A classic example for such system is the aqueous dispersion of Laponite ® particles, synthetic magnesium silicate particles with disk-like shape, diameter of 25-30 nm and thickness of 1 nm [13,14]. These particles have unique properties governed not only by their anisotropic shape, but also on the opposite charge present within the same particle in water – the surface of the particles is negatively charged due to the dissociation of $Na^+$, whereas the edges have positive charges at pH below 11 [14]. These unique properties allow gelation of Laponite dispersions, the rate of which depends strongly on both particles and salt concentrations [15].

An interesting example in which these two approaches can be potentially combined are the nonionic alkyl sucrose ester (SE) surfactants. Sucrose esters are biodegradable, non-toxic surfactants derived from the esterification of long-chain fatty acids (typically $C_{12}$-$C_{18}$) and sucrose [16,17]. Depending on the number of fatty acid residues attached to the sucrose unit, the hydrophilic-lipophilic properties can be finely tuned and varied from sucrose esters which are entirely water-soluble, when the monoesters predominates, to oil-soluble SEs, when di-, tri- and higher esters are predominantly present [17]. This variety allows wide utilization of these surfactants. Numerous studies demonstrate that sucrose esters are not only excellent candidates to replace the typical synthetic surfactants, but also have numerous beneficial properties, including anti-microbial, anti-tumor, anti-inflammatory and anti-oxidant activity [16-19]. Furthermore, they have been studied as additives enhancing the drug permeability [16,19].

In a recent study we demonstrated that even when ca. 20% diesters are present, along with 80% of sucrose monoesters in sucrose palmitate solution (ca. 80% $C_{16}$ and 20% $C_{18}$ fatty acid chains), these diesters do not completely dissolve into the water, nor include into mixed monoesters-diesters micelles. Instead, part of them phase-separate and form nanometer-sized particles [20]. The presence of particles makes these disperse systems better described as suspensions rather than solutions, therefore in the present study we will refer to them as dispersions. These particles co-exist with the slightly elongated monoester micelles at temperatures where the surfactant tails are in their frozen state. Upon heating, the tails undergo solid-to-liquid phase transition, which leads to reorganization of the sucrose ester molecules and formation of mixed wormlike micelles which cause significant increase of the viscosity of the solutions even at relatively low surfactant concentrations (2-5 wt. %) [20].

Furthermore, numerous studies have reported the presence of high negative charge in disperse systems containing SEs in absence of other ionizable molecular species [21-26]. Depending on the specific experimental conditions and the studied SE, the reported zeta potentials, $\zeta$, varied between ca. -71 mV and -31 mV at pH around 7 [21-25]. A decrease in the absolute value of the $\zeta$-potential has been reported in Ref. [22] upon decrease of the pH of the dispersions. Note that sucrose esters are nonionic surfactants, which by definition should be non-ionizable. Although the exact reason for this highly negative ζ-potentials has not been studied in details, there are three different explanations given in the literature: (1) the presence of residual non-esterified fatty acid remaining as impurity after the sucrose ester synthesis [21,22]; (2) the presence of free fatty acid produced due to chemical degradation (hydrolysis) in the system [22]; or (3) hydroxyl anions adsorption on the surface of supramolecular surfactant aggregates formed in the SE dispersions [23]. Further investigations are needed to determine which of these different explanations is really valid for SE systems.

The presence of negative ζ-potentials in SE dispersions in water suggests that their properties may be strongly affected by the addition of background electrolyte which will change the electrostatic interactions in the systems. Therefore, in the present study we aim to establish a relationship between the physicochemical properties of the SE dispersions and the added electrolyte concentration, and to explain the mechanism governing the observed changes. Furthermore, the role of the SE concentration, electrolyte type and pH value were also evaluated. The results demonstrate an alternative route for control of the rheological properties of sucrose esters containing systems which mechanistically differ from the presently known ones for nonionic surfactants.

## 2.      Materials and methods

### 2.1.      Materials

In the present study, we used sucrose fatty acid ester surfactant Ryoto sugar ester S-970, denoted in the text as S970 for simplicity, which was kindly provided by Mitsubishi Chemical Corporation. According to the information provided by the producer, it contains about 50% monoesters, and 50% di-, tri- and higher esters, with the diesters being the predominant fraction. About 70% of the fatty acid residues attached to the sucrose molecules are stearic ($C_{18}$), whereas the others are mainly palmitic ($C_{16}$). Assuming a chemical composition of randomly distributed $C_{16}$ and $C_{18}$ tails between monoesters, ME, and diesters, DE, (35 % $C_{18}$ME, 15 % $C_{16}$ME, 35 % $C_{18}$DE and 15 % $C_{16}$ME), the averaged molecular weight calculated based on this composition is $M_w \approx 729.4$ g/mol. Using the classic definition of the hydrophilic-lipophilic balance (HLB) provided by Griffin (HLB = 20 × $M_{hydrophilic\ part}/M_{total}$), the HLB of S970 can be estimated to be ≈ 10.8. Note that the HLB value provided by the producer is HLB ≈ 9, but it is derived using a different definition, HLB ≈ 20 × [monoester content, %]/100 [16,27].

To study the phase behavior of S970, we used various electrolytes, incl. sodium chloride (NaCl, purchased from Sigma), potassium chloride (KCl, Fluka), cesium chloride (CsCl, Valerus Bulgaria), magnesium chloride ($MgCl_2.6H_2O$, Sigma), calcium chloride ($CaCl_2.6H_2O$, Sigma), and aluminum chloride ($AlCl_3.6H_2O$, Sigma). All electrolytes had purity ≥ 99%. Furthermore, sodium hydroxide (NaOH, Merck, purity ≥ 99%), hydrochloric acid (HCl, Fluka, 37%) and stearic acid ($C_{18}Ac$, purity > 98%, TCI Chemicals) were also used in part of the experiments. All dispersions were prepared with deionized water purified by Elix 3 Module, Millipore USA.

### 2.2.    Experimental methods

### 2.2.1.  Dispersions preparation

The S970 dispersions were prepared following the procedure: the needed amount of surfactant powder was accurately weighted on an analytical balance. After that, the required amount of deionized water was added and the mixture was homogenized by stirring at a temperature of 77°C for about 30 minutes, until the surfactant become completely dissolved and slightly opalescent solution with low viscosity was obtained. Afterwards, the solution was cooled to 25°C and used for further experiments or for the preparation of samples with added electrolytes.

The samples containing salts were prepared at 25°C by mixing a concentrated S970 dispersion (usually 2.5 wt. % S970 dispersion was used for the preparation of samples containing 2 wt. % or less S970 final concentration, or 5 wt. % S970 dispersions for samples with 4 wt. % S970) with small amount of concentrated salt solution and the required amount of deionized water to achieve the desired salt/surfactant concentrations. Afterwards, the samples were homogenized by intensive hand shake or by stirring at 300 rpm for 3 minutes. The prepared samples were left to equilibrate at 25°C for at least one night until further experiments were performed with them.

The dilution approach was chosen as it allows much more precise addition of very small amounts of salts, which was needed for part of the samples. However, in some preliminary experiments, we also prepared samples by directly mixing the S970 powder, required salt and water, and then heated the dispersion to 77°C to homogenize it. Similar experimental results were achieved with these samples as well. The only significant difference which was observed was that the samples containing electrolyte concentrations above the threshold quantity for collapse of the gel, undergo phase separation faster compared to the samples which were prepared at 25°C.

### 2.2.2. Optical microscopy observations

All observations were performed with AxioImager.M2m microscope (Zeiss, Germany) in transmitted, cross-polarized white light or in reflected white light. For the experiments in transmitted light, we used a compensator plate placed after the sample and before the analyzer, which was oriented at a 45° angle with respect to both the polarizer and the analyzer to enhance the contrast of the sample. The sample was placed on a glass slide and covered with a microscope glass cover slip. For the experiments in reflected light, the same experimental set-up was used, but in this case after placing the sample over the glass slide, we introduced air bubbles into the sample using a needle attached to a syringe before covering the sample with the glass cover slip for observations. The thin aqueous films surrounding the bubbles were observed in reflected light, which allowed better visualization of the entities present within the studied solutions.

### 2.2.3. Rheological measurements

The rheological properties of the studied samples were characterized with a rotational rheometer Anton Paar, MCR-302e, using a cone and plate geometry (40 mm cone diameter, 1° cone angle, truncation gap 78 μm). To check the effect of possible wall slip over the obtained results with a cone-plate geometry, for part of the samples we also used a DHR-3 rheometer, TA Instruments. In this case parallel plates geometry was used (40 mm diameter of the upper plate, gap 300 μm). Fine sandpaper (grade P1500) was attached to both upper and lower plates before the measurements to prevent a possible wall-slip between the sample and the plates during the measurements. Two types of oscillatory shear rheological measurements were performed to characterize the viscoelastic properties of the samples: amplitude sweeps, where the oscillation amplitude was varied between 0.01 and 100% at a fixed frequency of 0.16 Hz, and frequency sweeps, in which the oscillation frequency was varied between 0.01 and 75 Hz at a fixed amplitude of 0.5%. The yield stress of a given sample was determined from the maximum of the elastic stress (equal to elastic modulus, $G$', multiplied by the absolute oscillation strain γ) vs. oscillation strain (γ) curve [28]. The viscosities were measured in flow ramp experiments, where the shear rate was logarithmically varied between 0.01 and 500 $s^{-1}$. All measurements were performed at 25°C. For all studied systems, we measured at least two independently prepared samples to ensure reproducibility of the obtained results.

### 2.2.4. Zeta-potential measurements

The zeta-potentials of the studied samples were conducted using a Zetasizer Nano ZS (Malvern Panalytical, UK). Samples with low ionic strength (I ≤ 9 mM) were analyzed in disposable folded capillary cells (DTS1070), while those with higher ionic strengths were measured using high-concentration cell ZEN1010. For each sample, at least three independent measurements were performed to ensure accuracy and reproducibility. All experiments were performed at 25°C temperature.

The Smoluchowski equation was used to covert the measured electrophoretic mobility, $U_e$, into a zeta potential, $\zeta$ [29]:

$$\zeta = \frac{U_e \eta}{\varepsilon \varepsilon_0},$$

(1)

where $\varepsilon$ is the dielectric constant for water solution 78.5, $\varepsilon_0$ is the permeability of vacuum, and $\eta$ is the viscosity of water which is 0.8872 mPa.s at 25°C. The equations used to calculate the surface potential and surface charge density are given in Section 3.5 below. It should be noted that the calculated $\zeta$-potential using the Smoluchowski equation tends to overestimate the surface charge density on the particle surfaces. This overestimation arises because this equation does not account for the slip boundary condition related to the hydrodynamic force on the particle surface. This point is further discussed in Section 3.5 below.

### 2.2.5. pH and conductivity measurements

The pH and electrical conductivity of the dispersions were determined using Fisher brand AB200 benchtop pH/conductivity meter at 25°C. The electrical conductivity measured for 2 wt. % S970 dispersion in absence of added electrolyte was ≈ 0.07 mS/cm. This conductivity is equivalent to the conductivity of ca. 0.45 mM NaCl solution (equation 2 from Ref. [30] was used for the conversion), which is negligible compared to the electrolyte concentrations studied within the current paper.

### 2.2.6. Laser diffraction

The size of the diester particles was measured using laser diffraction equipment Analysette 22 NanoTec (Fritsch GmbH, Germany). Measurements were performed between 0.005 and 40 μm and the detected signal was analyzed using the Mie scattering theory.

### 2.2.7. SAXS/WAXS measurements

The SAXS experiments were performed using the same procedure described in our previous study [20]. Briefly, Xeuss 3.0, Xenocs, France equipment was used with Eiger2 4 M detector (Dectris Ltd., Baden Deattwil, Switzerland) and CuKα X-ray source (λ = 0.154 nm). The samples were placed in special glass capillaries (WJM Glass, Germany) whose temperature was controlled using HFSX350 high-temperature stage, equipped with a T96 temperature controller and an LNP96 liquid nitrogen pump, all products of Linkam Scientific Instruments Ltd., UK. Sample-to-dectector distance of 750 mm was used and the acquisition time was set to 600 seconds. Further details are available in Ref. [20].

### 2.2.8. Cryo-TEM observation

The cryo-TEM observations were performed using a Gatan cryo-specimen holder at JEM2100, JEOL high-resolution transmission electron microscope. An acceleration voltage of 200 kV was used. Micrographs were recorded with the Gatan Orius SC1000 camera. The samples were prepared using Vitrobot system (FEI, USA) at 100% relative humidity and temperature of 25°C. Further details about the procedure can be found in Ref. [20].

## 3. Experimental results and discussion

### 3.1. Gels formation in presence of NaCl

The 2 wt. % S970 dispersion is a slightly opalescent, low-viscosity liquid with an apparent viscosity ≈ 0.005 Pa.s. A significant change in the macroscopic appearance of the sample was observed when prepared in the presence of NaCl, see **Figure 1a**. At low concentrations, c(NaCl) ≤ 4-5 mM, the dispersions remained freely flowing, low-viscosity liquids. However, as the NaCl concentration increased beyond 6 mM, a sharp rise in viscosity was observed, but the dispersions with c(NaCl) ≤ 8.5 mM were able to flow under the buoyancy force.

Samples containing 7-8.5 mM NaCl appeared as viscous liquids. Increasing the NaCl concentration to 9 mM induced the formation of non-flowing, gel-like samples, despite the surfactant concentration being just 2 wt. %. The lowest salt concentration required to achieve gel formation is referred to as critical concentration for gelation (c.c.g.) in this study. Note that although we will refer to the samples able to withstand their own weight under gravity as 'gels', these samples do not meet the strict rheological criteria for 'gel'. According to this definition, gels should exhibit a pronounced plateau in the moduli measured as a function of the frequency in oscillatory shear experiments [31]. The studied gel-like samples did not exhibit such behavior, although the elastic modulus exceeded the loss modulus across the entire tested frequency range (between 0.1 and 500 rad/s), no plateau was observed, see **Supplementary Figure S1c**.

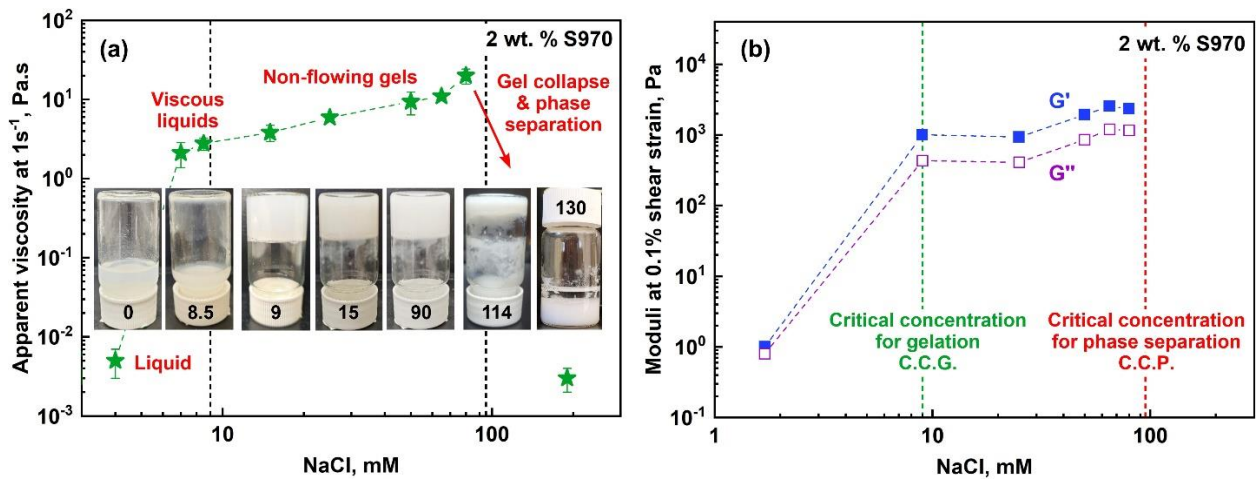

**<u>Figure 1.</u>** **Effect of NaCl concentration over the rheological behavior of 2 wt. % S970 dispersions.** (a) Apparent viscosity measured at 1 $s^{-1}$ (for c(NaCl) ≥ 7 mM; for 4 mM data at 100 $s^{-1}$ are presented due to the high error in the values measured at lower shear rates) as a function of the NaCl concentration. Inset: macroscopic pictures of the samples. The numbers denote the NaCl concentration in mM. (b) Storage (filled blue symbols) and loss (empty purple symbols) moduli measured at 0.1% shear strain in an amplitude sweep oscillatory experiments as a function of NaCl concentration. The experimental data shown in the figures are averaged of at least from two independently measured samples.

The apparent viscosity of the gel-like samples, measured at a shear rate of 0.01 $s^{-1}$, was ca. $10^3$ Pa.s, see **Supplementary Figure S1a**. This value is over five orders of magnitude higher than that of the corresponding salt-free dispersion. Furthermore, these gel-like samples, prepared with just 2 wt. % surfactant, demonstrated pronounced viscoelastic properties with the elastic modulus ($G$') exceeding the loss modulus ($G$'') at low shear strains, see **Figure 1b** and **Supplementary Figure S1b** as well.

Gel formation was observed up to a NaCl concentration of ≈ 95 ± 5 mM. Beyond this threshold, denoted as critical concentration for phase separation (c.c.p.), the viscosity began to decrease, the gels collapsed, and at NaCl concentrations exceeding ≈ 110 mM, phase separation occurred, producing a powder-like whitish layer which sedimented to the bottom of the container and a transparent, low-viscosity layer, see **Figure 1a**. Gels prepared at intermediate NaCl concentrations remained completely stable for over four months. Minimal water separation was observed in gels prepared at NaCl concentrations near the c.c.p. Notably, the original appearance of these samples could be fully restored through a simple heating and subsequent cooling.

### 3.2. Gelation mechanism

The observed phase behavior was unexpected, considering that S970 is a nonionic sucrose ester surfactant, whose properties are not expected to be affected by the addition of electrolytes [32]. To understand the mechanism underlying the observed transitions from low viscous samples to perturbed gels, gels, collapsed gels, and eventually phase-separated systems upon increase of the salt concentration in the dispersions, we first considered the arrangement of S970 molecules in the aqueous phase. According to the manufacturer, S970 surfactant contains about 50% sucrose monoesters and 50% di- and higher esters, with the diesters fraction being predominant. While the monoesters are easily soluble in water, the diesters are significantly more hydrophobic and cannot dissolve in water.

In our previous study, investigating the role of the monoesters-to-diesters ratio on the phase behavior of palmitic sucrose ester, we showed that even at a monoester-to-diesters ratio of 4:1, part of diesters could not be solubilized within the monoester micelles. Instead, they separated into individual nanoparticles with size about 50 nm and aggregates of nanoparticles [20]. A similar process was expected for the S970 dispersions, which have a significantly higher monoester-to-diesters ratio ($\approx$ 1:1 instead of 4:1). Note that the critical packing parameter, $CPP = v_0/(a_0 l_0)$, where $v_0$ is the volume of the surfactant tail, $a_0$ is the optimal surface area of the surfactant molecule, and $l_0$ is the length of the surfactant tail [33], for the $C_{16}$ and $C_{18}$ sucrose monoesters is $\approx$ 0.49, whereas it is $\approx$ 0.98 for the diester molecules. In this calculation, we have used $a_0 \approx 45.6$ Å$^2$ [34], $l_0 \approx 20.52$ Å and 23.05 Å for $C_{16}$ and $C_{18}$ tails, respectively, and $v_0 \approx 458.1$ Å$^2$ and 511.9 Å$^2$ for $C_{16}$ and $C_{18}$ monoesters [34,35]. Therefore, the diester molecules will preferably arrange in lamellar structures with zero mean curvature.

The turbidity of the S970 dispersions, in the absence of added salt, directly indicates the presence of supramolecular aggregates capable of scattering visible light, see insets in **Figure 1a** and **Figure 2b**. Laser diffraction measurements of the S970 dispersions revealed objects with a mean Sauter diameter, $d_{32} \approx 85 \pm 5$ nm and mean volume diameter, $d_{V50} = 130$ nm. Additionally, the presence of micelles and a small number of larger particle aggregates was observed, see **Figure 2a** and **Supplementary Figure S2a**. The presence of relatively monodisperse diester particles coexisting with slightly elongated micelles at 25°C, was further confirmed by the SAXS spectra measured for S970 dispersions, see **Supplementary Figure S2b**. Similarly to the palmitic surfactant, studied in Ref. [20], the peaks showing diester particles in the S970 dispersion disappeared only when the temperature was increased above the phase transition temperature for the S970 surfactant ($T \approx 48°C$), which was accompanied by changes in the micellar shape and size. Note that the disappearance of these particles is also expected to affect significantly the presently studied gel formation. Future study of this effect is needed to describe the temperature effect over the presently investigated phenomenon.

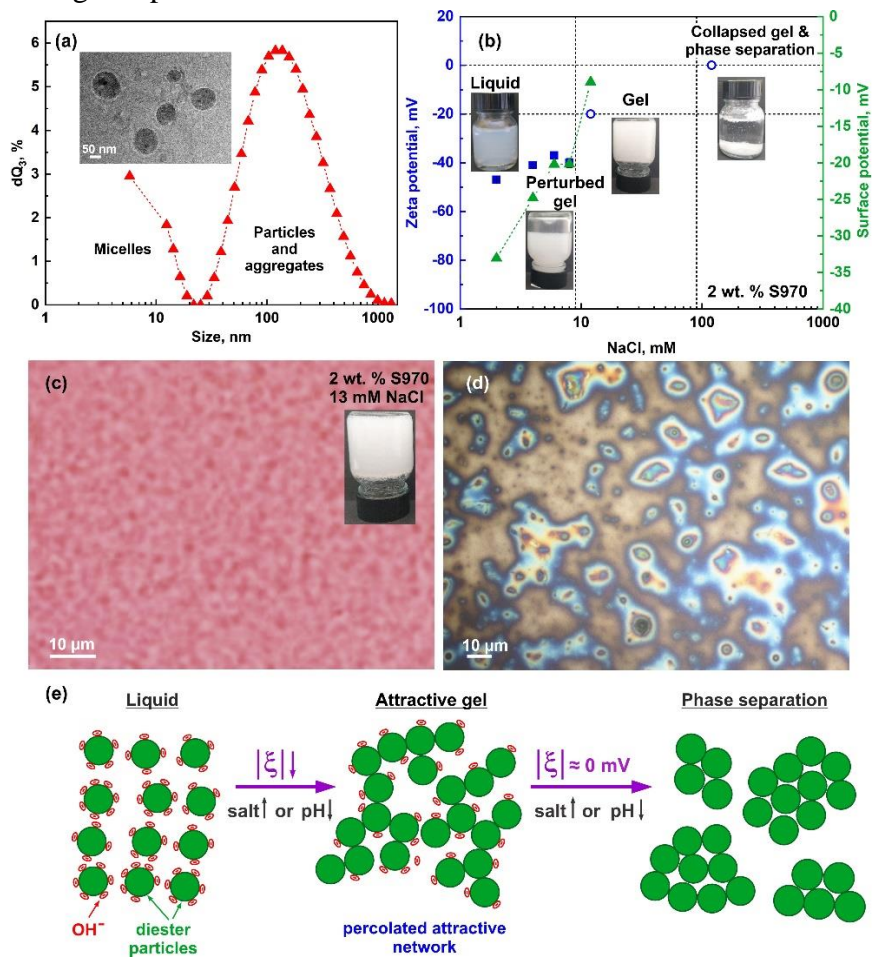

**Figure 2.** Mechanism of gel formation for S970 dispersions upon addition of electrolyte. (a) Histogram of particle size (diameter) distribution by volume measured with laser diffraction instrument for 2 wt. % S970 dispersion averaged from three measurements. Inset: Cryo-TEM picture showing the diester particles. (b) $\zeta$-potential (blue symbols, left axis) and calculated surface potential by eq. (3) (green triangles, right axis) as a function of the NaCl concentration averaged

from three measurements. Inset: illustrative pictures of the samples in different regions. (c) Optical microscopy picture of 2 wt. % S970 + 13 mM NaCl sample. This sample is a non-flowing gel. (d) Optical microscopy picture in reflected light showing the inhomogeneous thin film thickness due to the entrapped particles within the film. (e) Schematic representation of the phase behavior experienced by S970 dispersions upon addition of electrolyte. The micelles formed by the monoesters are disregarded for simplicity, as well as the bigger aggregates which are initially present in the sample. The schematics illustrates that the decrease of the absolute value of the zeta potential allows formation of percolated space-spanning network of particles which results in formation of macroscopic gel until the charges are fully screened, where the gel collapses and phase separation occurs.

The observed change in the rheological properties of the S970 dispersions upon the addition of NaCl, **Figure 1**, suggests the presence of electrostatic interactions. As explained in the introduction, although the sucrose ester surfactants are nonionic, various studies report highly negative zeta potentials in such systems [21-26]. Therefore, next we measured the zeta potential of 2 wt. % S970 dispersions at different salt concentrations, see **Figure 2b**. In the absence of added electrolyte (ionic strength of 0.45 mM due to presence of salt in the used compound), the measured value was $\zeta \approx$ -55 ± 5 mV. This result is in good agreement with the results reported in Ref. [25], where the authors measured a zeta potential of -57 mV for nanoemulsions of cetearyl ethylhexanoate and isopropyl myristate oils emulsified in 2.5 wt. % S970.

The increase of the NaCl concentration in 2 wt. % S970 dispersions resulted in a decrease in the absolute zeta potential measured. For the sample containing 10 mM NaCl, which is just above the critical concentration needed for the formation of a completely gelled sample in the presence of 2 wt. % S970, the measured zeta potential was $\approx$ -20 mV. Note that this zeta-potential was measured shortly after the sample was prepared, before it became a gel, which required several hours of storage at 25°C. Additionally, the zeta potential of the freely flowing, transparent liquid phase, separated by filtration from the 130 mM NaCl-containing sample after its phase separation, was -5 ± 5 mV. Similar zeta potential from salt concentration dependence was obtained with 0.5 wt. % S970 dispersions, see **Supplementary Figure S3**. Gelation was not observed at this surfactant concentration, which allowed measurements to be performed throughout the entire NaCl concentration range.

These results demonstrate that the observed gelation is mainly governed by electrostatic interactions between the diester particles (aggregates) in dispersion, see the schematic mechanism of phase behavior shown in **Figure 2e** and optical microscopy images showing the gel structure in **Figure 2c,d**. At low NaCl concentrations, the electrostatic repulsion between the particles is too strong, preventing them from coming into close contact with one another. The addition of salt, however, partially screens the charges on the particle (or aggregate) surfaces, as evidenced by the decrease in the zeta potential. This allows for a sol-gel transition. The gel state exists within a certain NaCl concentration range, where the electrostatic repulsion is weak enough to enable particle arrangement into a network that occupies the entire sample volume, while still preventing all of them from coming into too close contact, -20 mV $\leq \zeta <$ 0 mV. When the NaCl concentration becomes high enough to completely screen the charges, particles begin to aggregate irreversibly, causing the gel to collapse and begin to flow. Upon further increase of NaCl concentration, or upon prolonged storage, phase separation occurs, with particles settling at the bottom of the container, see **Figure 2b**. Therefore, the phenomenon of gel formation at very low surfactant and salt concentrations results from the in-situ formation of a network of interacting particles. Although these particles consist of surfactant molecules, their size (130 nm in diameter, see **Figure 2a**) makes them similar to inorganic solid nanoparticles, for which similar phenomena have been previously observed, as described in Section 3.6. Therefore, our system makes a connection between surface science and colloidal science. Since such effects are not expected for individual surfactant molecules, we discuss our results from the colloidal science perspective in further details in Section 3.5 below. We note that the currently employed concentrations at which gels are observed to form are significantly higher than the critical micellar concentration for S970 surfactant (found to be ca. $5 \times 10^{-4}$ wt. % in water, 5 mM NaCl and 50 mM NaCl, data not shown), demonstrating that the observed effects are governed by the volume fraction of the diester particles, rather than the micelles formation.

The gels formed by interactions between suspended particles in dispersions are usually classified as either repulsive Wigner glasses or attractive gels [15]. In the repulsive Wigner glass state, particles are disconnected, and the gel is stabilized by the electrostatic repulsions. Conversely, the attractive gel state also results in macroscopic gel formation, but the 'particles form a percolated space-spanning network' [15].

For 1-4 wt. % S970 dispersions, the gels obtained upon addition of small amount of electrolyte were found to be attractive gels. Dispersions prepared without added salts, which had the highest electrostatic repulsion between the particles, did not undergo sol-gel transition even after several months of storage (the higher the electrostatic repulsion, the faster the phase transition should be for the Wigner glass state). In contrast, the sol-gel transition was significantly accelerated with increase of the salt concentration (within the NaCl gel concentration range). This observation is in good agreement with the DLVO theory, which suggests that the structure formation rates increase with salt concentration, leading also to a decrease in the Debye screening length.

### 3.3. Origin of the negative zeta potential in nonionic sucrose esters systems

The experiments described in Section 3.2 revealed that the gelation observed in S970 dispersions is driven by an attractive interaction between the diester particles (aggregates), which form a percolated, space-spanning network when the electrostatic repulsion between them is partially screened, see **Figure 2**. However, since sucrose esters are nonionic surfactant, the origin of the electrostatic repulsion remains unclear. As discussed in the introduction, two main hypotheses have been proposed in the literature to explain the observed negative charge: the presence of free fatty acid molecules (either as residual impurities after SE synthesis or due to ester hydrolysis) [21,22], or the adsorption of hydroxyl anions on the particle surfaces [23]. To determine which of these explanations applies to the currently investigated system, we performed additional experiments.

First, we investigated the potential effect of the presence of free fatty acids in the dispersions. According to the analysis sheet provided by the producer, the currently used S970 surfactant contains no more than 0.9 % free fatty acids. Given that the surfactant concentration explored is 2 wt. %, this corresponds to $\leq 0.018$ wt. % free fatty acid in the studied systems. Since the majority of alkyl chains attached to the surfactant molecules consists of 18 C-atoms, we investigated the phase behavior of dispersions with externally added stearic acid ($C_{18}Ac$). We note that the pKa values of long-chain fatty acids are around 8.5-10 (8.7 for palmitic acid, 9.85 for the oleic acid, and 10.15 for stearic acid [36]). Therefore, even if there are some fatty acids present as impurities in the S970 surfactant they would be in their neutral stage at the pH values studied which varied between ca. 5.5 and 7.5 depending on the surfactant concentration, and on the pH of the initial deionized water used for the preparation of the samples. Nevertheless, to completely exclude the possible effect of free fatty acids, we conducted experiments in which stearic acid was purposely added to S970 dispersions.

The zeta potential measurements for 2 wt. % S970 dispersions with added $C_{18}Ac$ (ranging from 0.04% to 0.3% concentration) were similar to those without added stearic acid, $\zeta \approx$ - 62 $\pm$ 5 mV for the samples containing acid and ζ ≈ - 55 ± 5 mV for the samples without externally added acid. Furthermore, the NaCl concentration needed for gelation of the sample with 0.16 wt. % C$_{18}$Ac (which is about 10 times higher concentration of acid compared to the potentially fatty acid concentration in the original sample) was the same as that required for the gelation of samples without added acid, c(NaCl) ≈ 9 mM. These results demonstrate that the highly negative zeta potential observed in the S970 dispersions and the gelation phenomenon cannot be explained by the presence of a negligibly small amount of free fatty acid. Any influence from the presence of free fatty acids on the observed negative zeta potential (if present at all) is of secondary order.

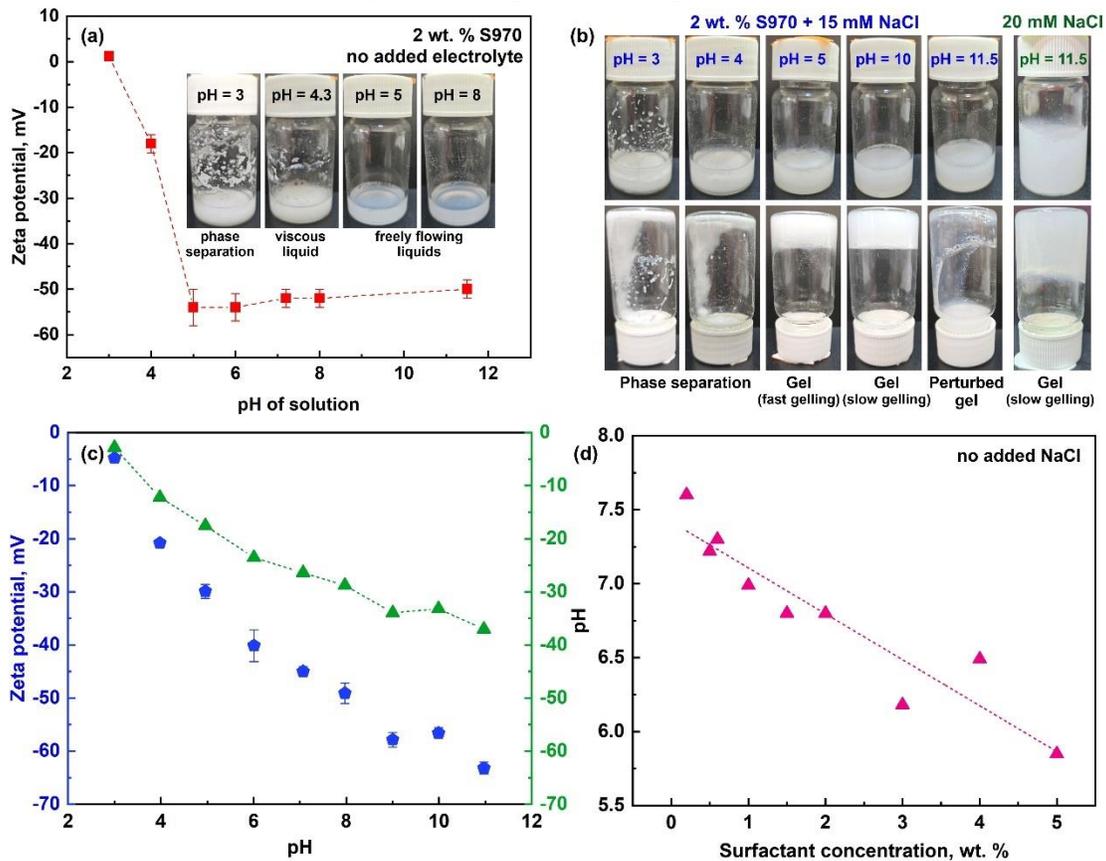

**Figure 3.** **Model experiments showing the origin of the negative zeta potential in S970 dispersions.** (a) Effect of pH for phase behavior of 2 wt. % S970 dispersions. Samples prepared without added NaCl. A sharp increase in the zeta potential is observed for samples with pH < 4. This also changes the phase behavior of the samples from freely flowing liquids to viscous liquids at pH ≈ 4.5. At pH = 3, the ζ ≈ 0 mV, which leads to aggregation of the diester particles and phase separation of the dispersion. The data for ζ-potentials are averaged from three measurements (b) Macroscopic pictures of samples prepared with 15 mM NaCl (blue labels) or 20 mM (green label, last pictures) at different pH values. (c) Zeta potential (blue pentagons, left axis) and calculated surface potential by using eq. (3) (green triangles, right axis) dependence on pH for 0.2 wt. % S970 dispersions prepared at a constant ionic strength $I$ = 5 mM. The NaOH or HCl added to adjust the pH is accounted towards the final ionic strength of the dispersion and the necessary amount of NaCl dispersion is added to achieve the equal ionic strength for all samples. The data for $\zeta$-potentials are averaged from three measurements. (d) pH dependence on S970 concentration. The deionized water used for dispersions preparation had pH $\approx$ 8. The data are averaged from three measurements

Another possible reason for the highly negative zeta potential observed in S970 dispersions is the adsorption of hydroxyl anions on the diester particles (aggregates). Note that hydroxyl ions are always present in aqueous solutions due to the autoprotolysis of water. Their concentration depends strongly on the pH of the solution. Previous studies have attributed the presence of a negative zeta potential in the range of -50 to -60 mV, at an ionic strength of 1 mM and pH = 6, to the specific spontaneous adsorption of hydroxyl anions at the oil-water interface [37].

To assess the effect of pH on the phase behavior of the S970 dispersions in the presence of 15 mM NaCl and in the absence of added electrolyte, the following experiments were conducted. A stock dispersion of 2.05 wt. % S970 was prepared and allowed to cooled to 25°C. Samples with pH values ranging from 3 to 11.5 were prepared by adding small amounts of concentrated HCl or NaOH, along with the required amount of water to achieve a 2 wt. % S970 concentration. For the experiments with NaCl, the same procedure was followed, but we also added a small quantity of 1M NaCl solution to adjust the NaCl concentration in the samples to 15 mM. The results from these experiments are summarized in **Figure 3a,b**.

For samples without added salt, a slight increase in viscosity was observed at pH = 4.5. The sample prepared at pH = 4.3 was almost gel, whereas when the pH was decreased to 4, the sample became highly viscous collapsed gel. Further decrease in pH to 3 resulted in aggregation of the diester particles and phase separation of the sample. Accordingly, the zeta potential of the transparent layer of this sample was found to be around 0 mV.

A different phase behavior was determined for samples prepared with 2 wt. % S970 and 15 mM NaCl at varying pH values, see **Figure 3b**. Gels formed when the pH ranged from 5 to 10. However, gelation occurred much faster (within several minutes) at pH = 5 due to the lower concentration of hydroxyl anions, compared to the gelation in samples with pH values between 6 and 10, where several hours were needed to obtain non-flowing gel-like samples. At pH values of 3 and 4, the samples rapidly underwent phase separation. The sample prepared at pH 4.5 appeared as collapsed gel with small amount of separated water. In contrast, the sample prepared at pH =

11.5 appeared as highly viscous but flowing gel, likely due to the significantly increased hydroxyl anions concentration, which made the 15 mM NaCl concentration insufficient to effectively screen the electrostatic repulsion between particles and allow formation of attractive gel.

This hypothesis was validated through zeta potential measurements conducted for 0.2 wt. % S970 dispersions over a pH range of 3 to 11, see **Figure 3c**. A low surfactant concentration was chosen to eliminate any potential viscosity effects on the zeta potential measurements, as these dispersions remained with low viscosity across the entire pH range studied. Furthermore, a constant ionic strength, $I$ = 5 mM, was maintained for all samples, to ensure consistent positioning of the diester particles relative to the plane where the electrophoretic mobility is measured and zeta potential is calculated (note that this is not the case for results shown in **Figure 3a**). As expected, an increase in pH led to a significant rise in zeta potential and vice versa, **Figure 3c**. This indicates that the adsorption of hydroxyl anions on the particle surface is pH-dependent, influencing the minimum salt concentration required to reduce the zeta potential and achieve the percolation threshold. In agreement, when the NaCl concentration for a 2 wt. % S970 sample at pH 11.5 was increased from 15 mM to 20 mM, the formation of a non-flowing gel was observed, see the last pictures in **Figure 3b**.

Further evidence for hydroxyl anion adsorption as the cause of the negative zeta potential was observed when preparing samples with varying S970 concentrations. Dissolving 5 wt. % S970 in deionized water with an initial pH of ≈ 7.7 resulted in a decrease in pH to 5.9. Upon dilution with the same deionized water, the pH of the dispersions increased progressively as the S970 concentration decreased: pH ≈ 6.5 for 4 wt. % S970, pH ≈ 6.8 for 1.5-2 wt. % S970, and pH ≈ 7.6 for 0.2 wt. % S970, see **Figure 3d**. These results clearly indicate that the increase of the diester particles concentration leads to increase of the number of hydroxyl anions adsorbed on their surface, leading to an increase in protons concentration and a decrease in dispersion pH.

The obtained results demonstrate that the negative zeta potential in S970 dispersions results from hydroxyl anion adsorption on the surface of diesters particles (aggregates). Effective charge screening can be obtained by adding a background electrolyte or by decreasing the pH. Both approaches lead to the formation of highly viscous and non-flowing samples when the zeta potential falls between -20 mV and 0 mV. Next, we investigated the effect of surfactant concentration on the observed phase behavior.

### 3.4. Role of surfactant concentration for samples gelation. Viscoelastic properties of the gels.

The increase of surfactant concentration increases the volume fraction of insoluble particles, whose interactions are responsible for gel formation. To explore the role of surfactant concentration in the gelation process, experiments were performed at a fixed NaCl concentration of 34 mM. Based on the initial experiments performed with 2 wt. % S970, this salt concentration was found to be near the center of the gelation region, see **Figures 1a** and **2b**. The results, shown in **Figure 4a**, reveal an increase in apparent viscosity when the surfactant concentration increases. All samples prepared with S970 concentrations exceeding 1.3 wt. % appeared as non-flowing gels. The apparent viscosities of the samples containing 1.4 and 2 wt. % S970 were similar, $\eta_{1s-1} \approx 5$ Pa.s. A slight increase was observed when the surfactant concentration was doubled in the presence of 34 mM NaCl, $\eta_{1s-1} \approx 13$ Pa.s for 4 wt. % S970.

To investigate the effect of surfactant concentration on the minimum salt concentration required for gelation, we prepared samples containing 9 mM NaCl, identified as the threshold salt concentration for gel formation in 2 wt. % systems, and systematically varied the S970 concentration. Gels formed successfully in samples containing between 2 and 4 wt. % S970, though gelation kinetics varied significantly. One hour after its preparation, the sample with 4 wt. % surfactant significantly increased its viscosity becoming a perturbed gel. About 90 minutes were needed for this sample to become completely non-flowing gel. Longer time was needed for the development of a gel network within the samples containing lower surfactant concentration; for instance, a 2 wt. % S970 sample developed a gel network over $\approx 16$ hours. Samples with 1.32 wt. % S970 and 1.5 wt. % S970 exhibited significant viscosity increase after overnight storage but did not form non-flowing gels, even over extended periods. Their yield stresses were measured to be $\approx 4.3$ and 6.5 Pa, respectively. When NaCl concentration was increased to 10 mM, the yield stress for 1.5 wt. % sample increased to $\approx 7.5$ Pa and this sample became a non-flowing gel as well. This small increase in the c.c.g. threshold concentration at different S970 contents aligns with the pH-dependent gelation mechanism proposed. As illustrated in **Figure 3d**, higher S970 concentration results in lowed pH values. Consequently, the hydroxyl anion concentration in the 1.5 wt. % sample will be higher than that in 2 wt. % sample, making the c.c.g. for 1.5 wt. % sample slightly higher. In agreement with this reasoning, the critical concentration for gelation, c.c.g. of 4 wt. % S970 dispersions was determined to be ≈ 6 mM NaCl. However, it is noteworthy that the gel formation required about two days for this sample.

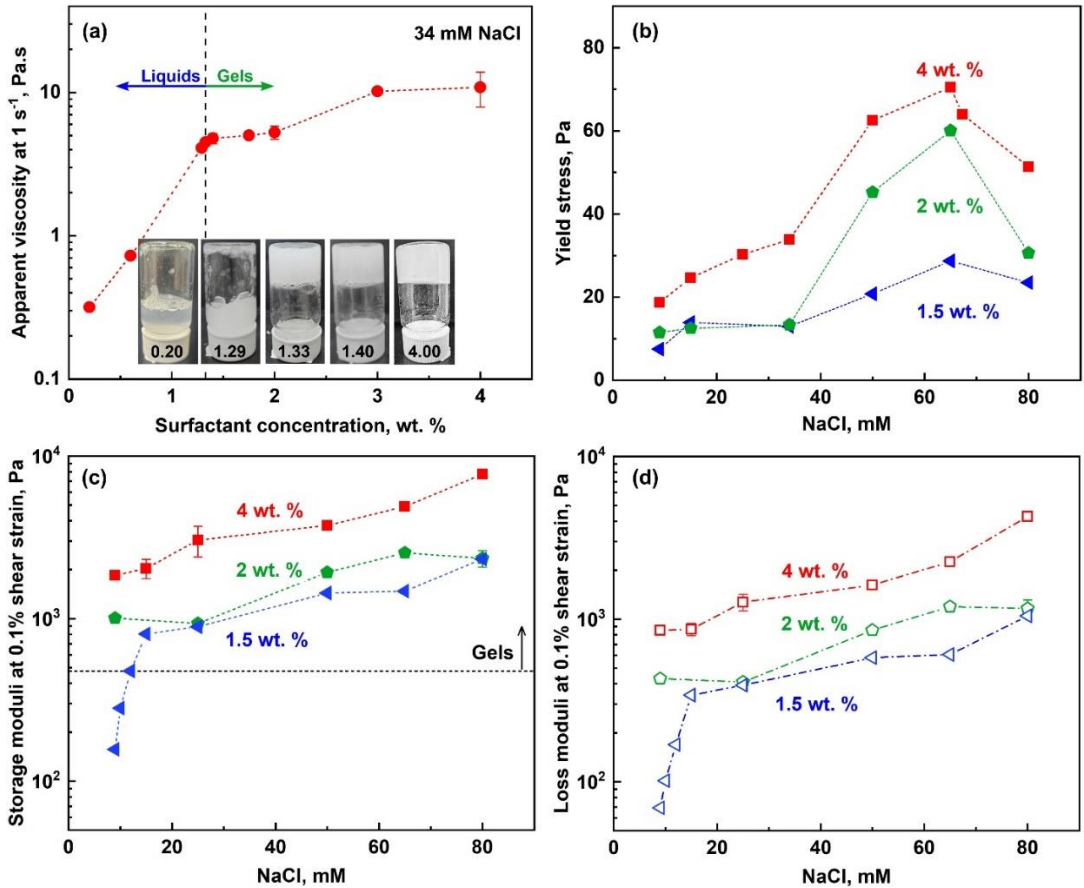

**Figure 4.** **Effect of surfactant concentration.** (a) Apparent viscosity measured at 1 s$^{-1}$ for S970 samples with 34 mM NaCl and different surfactant concentrations. The samples containing more than 1.3 wt. % surfactant appear as non-flowing gels. Inset: pictures of selected samples. The numbers denote the S970 concentration in wt. %. (b) Yield stress dependence on NaCl and S970 concentration. (c,d) Storage (filled symbols, c) and loss (empty symbols, d) moduli measured for different surfactant and salt concentrations: 1.5 wt. % - blue triangles; 2 wt. % - green pentagons; 4 wt. % - red squares. The experimental data shown in the figures are averaged from two measurements.

The dependence of viscoelastic properties on surfactant and salt concentrations was investigated through oscillatory rheological experiments, see **Figure 4b-d** and **Supplementary Figure S4**. For all samples, the elastic modulus dominated the viscous modulus until the yield stress of the sample was reached. Beyond this point, the samples began to flow, and shortly after that the viscous modulus become higher than the elastic modulus. Both salt and surfactant concentrations significantly influenced the moduli. Increase of surfactant concentration led to a notable rise in both the moduli and yield stress. A similar trend was observed with increasing salt concentration, but only within the gel concentration range. Once the salt concentration exceeded the gelation limit, phase separation occurred and the rheological properties of the samples could not be correctly determined anymore.

Yield stresses, $\tau_y$, shown in **Figure 4b**, were determined using the maximum in the elastic stress (G'$\gamma$) vs. shear strain ($\gamma$) curve obtained in amplitude sweep measurements performed with cone-and-plate geometry (data shown in **Supplementary Figure S4a**). To check for a potential wall slip during these measurements, additional tests were performed using parallel plates geometry with attached sandpaper to both plates and with sandblasted parallel plates. A comparison between the obtained curves is presented in **Supplementary Figure S4c**. As seen from it, when sandpaper is not used, the sample begin to flow at lower strain values due to the wall slip. However, the general trend that the increase of surfactant or salt concentration leads to increase of the yield stress remained consistent. Therefore, we chose to perform all experiments with the cone-and-plate geometry to facilitate direct comparisons between the rheological properties determined for flowing and gel-like samples (under constant shear rate applied to the whole sample). It is important to note, however, that the absolute yield stress values reported may be slightly underestimated due to wall slip effects.

The hysteresis in the rheological response of the prepared samples was investigated by performing experiments with 4 wt. % S970 + 50 mM NaCl as follows: the initial measurement was performed by increasing the shear rate from 0.01 to 100 s$^{-1}$, followed immediately by a decrease in shear rate from 100 s$^{-1}$ down to 0.01 s$^{-1}$. As shown in **Supplementary Figure S5**, the apparent viscosity data obtained when the shear rate decreased were slightly lower compared to the data from the increasing shear rate (see empty and full red points in **Supplementary Figure S5**). This difference indicates that the gel structure is perturbated by shearing at high shear rates. To determine the time required for the structure to rebuild, a new cycle of increasing and decreasing shear rate was performed after storing the sample for 5 min without shearing it (blue symbols in **Supplementary Figure S5**). As observed, the structure nearly rebuilt during this storage period, although the results at low shear rates were still slightly lower. Storing the sample for 30 min without shearing led to complete restoration of the gel structure, as can be seen from the data shown with green symbols in **Supplementary Figure S5**. These experiments clearly demonstrate that the formed gels can restore their structure over a relatively short period of time.

### 3.5. Role of electrolyte type

Results obtained solely with NaCl have been presented so far. The electrostatic screening mechanism identified as governing the observed phase behavior suggests that similar phenomena should occur in the presence of other electrolytes as well. As already shown, the role of the electrolyte is to reduce the electrostatic repulsion between the diester particles, enabling the formation of a space-spanning percolated network.

To investigate the phase behavior of S970 sucrose ester surfactant with different electrolytes, further experiments were performed using monovalent (KCl and CsCl), divalent ($MgCl_2$ and $CaCl_2$), and trivalent ($AlCl_3$) salts. These salts were selected to evaluate the effect of ion valence on the critical concentration for gelation, c.c.g., and the critical concentration for phase separation at which the gel collapses due to the complete suppression of electrostatic repulsion, c.c.p. Ion valence is well known to significantly affect the ability of counterions to screen electrostatic repulsion [32]. According to the Schulze-Hardy rule, the critical coagulation concentration (c.c.c.) dependency on ion valence, $z$, can be expressed as c.c.c. $\propto z^{-n}$, where the exponent $n$ usually varies between 2 and 6, depending on the specific colloidal system [38-40]. Exponent equal to 6 is expected for very high charge density on colloidal particles, whereas for poorly charged particles and low surface potentials, $n = 2$ is obtained based on the Debye-Hückel theory [41].

The results with monovalent salts revealed that the c.c.g. for 2 wt. % S970 was always 9 mM (ionic strength, $I = 9$ mM) independently on the cation type ($Cs^+$, $K^+$ or $Na^+$), see **Figure 5** and **Table 1**. The samples prepared with 8 mM monovalent salts appeared as viscous liquids, which approach the percolation threshold concentration but yet remained slightly below it. Significantly lower amount of divalent salt was needed to obtain non-flowing gel-like samples. The c.c.g. determined for $CaCl_2$ and $MgCl_2$ at 2 wt. % S970 was $\approx 1.5$ mM (1.45 mM for $CaCl_2$, $I = 4.35$ mM and 1.5 mM for $MgCl_2$, $I = 4.5$ mM). For the trivalent $AlCl_3$, however, stable gels were not prepared at 2 wt. % S970 concentration, see **Supplementary Figure S6**. The turbidity of the samples prepared with 0.1 mM $\leq AlCl_3 \leq 0.4$ mM progressively increased, but they remained as freely flowing low viscosity liquids. Further increase of the $AlCl_3$ concentration led to significant increase of the viscosity and partial formation of chunks of highly viscous parts within the sample (for $AlCl_3$ between 0.45 and 0.65 mM, corresponding to $I$ between 2.7 mM and 3.9 mM), but completely non-flowing gels could not be prepared. Instead the increase of $AlCl_3$ concentration to 0.7 mM ($I = 4.2$ mM) or above led to phase separation of the samples and surfactant precipitation.

**Table 1.** Critical electrolyte concentration for gelation (c.c.g.), critical ionic strength for gelation (CISG), critical electrolyte concentration for phase separation (c.c.p.), critical ionic strength for precipitation (CISP) for different electrolytes studied. The standard deviations show the resolution at which the different concentrations were studied. For each electrolyte, at least three independently prepared samples with each concentration around the critical one were prepared and evaluated.

| | Ion | Hydrated ion radius, nm [32] | Molar Gibbs energy of ion hydration, kJ/mol [42] | Hydration number (± 1) [32] | 2 wt. % S970 | | 4 wt. % |
| --- | --- | --- | --- | --- | --- | --- | --- |
| | | | | | c.c.g., mM (CISG, mM) | c.c.p., mM (CISP, mM) | c.c.g., mM (CISG, mM) |
| **CsCl** | $Cs^+$ | 0.33 | -250 | 1 | 9 ± 0.5 (9 ± 0.5) | 75 ± 5 (75 ± 5) | n.s. |
| **KCl** | $K^+$ | 0.33 | -295 | 3 | 9 ± 0.5 (9 ± 0.5) | 82.5 ± 2.5 (82.5 ± 2.5) | n.s. |
| **NaCl** | $Na^+$ | 0.36 | -365 | 4 | 9 ± 0.5 (9 ± 0.5) | 95 ± 5 (95 ± 5) | 6 ± 1 (6 ± 0.5) |
| **CaCl₂** | $Ca^{2+}$ | 0.41 | -1505 | 6 | 1.45 ± 0.05 (4.35 ± 0.15) | 9 ± 0.5 (27 ± 1.5) | 1.2 ± 0.10 (3.6 ± 0.5) |
| **MgCl₂** | $Mg^{2+}$ | 0.43 | -1830 | 6 | 1.50 ± 0.02 (4.5 ± 0.15) | 25 ± 0.5 (75 ± 1.5) | 1.4 ± 0.05 (4.2 ± 0.5) |
| **AlCl₃** | $Al^{3+}$ | 0.48 | -4525 | 6 | 0.45* (2.7) | 0.65* (3.9) | 0.7 ± 0.05 (4.2 ± 0.5) |

The "*" symbols included for 2 wt. % S970 with $AlCl_3$ denotes that these samples are not real gels. "n.s." = not precisely studied.

This result was attributed to the fact that the absolute number of $Al^{3+}$ ions present in dispersion is too low to allow formation of homogeneous gel. Assuming that the average molecular weight of the diester molecules is about 858 g/mol (30% $C_{16}$ chains and 70% $C_{18}$ chains) and that the surfactant contains ca. 50% diesters, it can be estimated that the ratio between $Al^{3+}$ ions and S970 diester molecules varies between 0.039 and 0.056 for $AlCl_3$ concentrations varied from 0.45 to 0.65 mM and 2 wt. % surfactant, i.e. for each ca. 18-26 S970 molecules there is a single $Al^{3+}$ ion in the dispersion. When the same estimate is made for the divalent salts, the calculated ions to surfactant molecules ratio within the gelation region is significantly smaller, between 0.5 and 8, whereas for the monovalent salts for each surfactant molecules there are between 1.3 and 8 counterions.

To check whether the increase of S970 content will allow the formation of non-flowing gel-like samples with AlCl₃, experiments were performed with 4 wt. % S970. Gels were successfully prepared with $0.7 – 1$ mM AlCl₃ ($I = 4.2\text{-}6$ mM) at this surfactant concentration. Similarly, to the general phase behavior exhibited by the samples prepared with monovalent and divalent ions at 2 wt. % surfactant concentration, at lower c(AlCl₃) $\leq 0.5$ mM ($I = 3.0$ mM), the samples appeared as freely flowing liquids; at 0.6 mM AlCl₃ ($I = 3.6$ mM) the sample increased its viscosity and appeared as perturbed gel; and at c(AlCl₃) $> 1$ mM ($I = 6$ mM), a phase separation was observed.

The c.c.g. determined for monovalent NaCl and divalent CaCl₂, and MgCl₂ at 4 wt. % S970 concentration were 6 mM ($I = 6$ mM), 1.2 mM ($I = 3.6$ mM) and 1.4 mM ($I = 4.2$ mM), respectively, **Table 1**. These concentrations are slightly lower compared to c.c.g. determined for 2 wt. % S970, most probably due to the lower pH of the 4 wt. % sample, see **Figure 3d**, which also determines the amount of hydroxyl anions adsorbed on the particles surface and the surface charge density of the in-situ formed diester particles.

The comparison between the critical concentrations for gelation of 2 wt. % S970 samples observed with monovalent, divalent and trivalent salts studied, showed that c.c.g. $\propto z^{-2.7}$, see **Figure 5a**. For the 4 wt. % S970, the exponent was slightly smaller, $n = 2$ for AlCl₃ and $n \approx 2.2$ for the divalent ions. These values for the exponent suggest a behavior closer to this for low surface charge density (low-potential) materials. Note that this result is reasonable considering that the charges come from the adsorption of the hydroxyl anions on the diester particles surface. Such behavior is also seen for carboxylic acid latex particles, for example, where usually $n = 2$ dependence is observed rather than a behavior of the high surface charge materials, where $n = 6$ is expected (for example for platinum sol) [38]. We note that exponent close to 3 has been previously observed for other colloidal particles, for example – amphoteric latex particles [38] and {Mo₇₂Fe₃₀} clusters [43].

The critical electrolyte concentration for gelation determined in the present study is relatively low, and the measured $\zeta$-potentials before gelation are also low. This suggests that the Debye-Hückel approximation of the DLVO theory can be used to determine the surface charge density at the critical ionic strength for gelation, using the expression derived in Ref. [36]:

$$CISG = \left(\frac{9}{8\pi \exp(2)}\right)^{1/3} \frac{1}{\lambda_B} \left(\frac{\sigma^2}{\varepsilon \varepsilon_0 A_H}\right)^{2/3}, \tag{2}$$

where CISG is the critical ionic strength for gelation, $\lambda_B$ is a Bjerrum length, which measures electrostatic interactions and has a value of 0.71 nm for interactions in water at 25°C, $\sigma$ is the surface charge density, $\varepsilon$ is the dielectric constant of water (78.5 at 25°C), and $A_H$ is the Hamaker constant, assumed to be $\approx 5\times10^{-21}$ J, as calculated for hexadecane-water-hexedacane in Ref. [30]. Using eq. (2), we calculated the magnitude of surface charge density at the gelation point to be 2.0 mC/m$^2$ for monovalent Na$^+$, K$^+$ and Cs$^+$, 1.2 mC/m$^2$ for divalent Mg$^{2+}$ and Ca$^{2+}$, and 0.8 mC/m$^2$ for AlCl$_3$.

To compare the determined charge densities from the CISG with those estimated from the measured $\zeta$-potential, we used the equation derived by Gopmandal et al. which relates the estimated $\zeta$-potential (determined on the base of Smoluchowski equation, eq. 1 above) with the surface potential, $\psi_0$, on the particle surface [44]:

$$\psi_0 = \zeta \frac{1}{\left(1 + 2\exp(\kappa a)E_5(\kappa a) - 5\exp(\kappa a)E_7(\kappa a) + \frac{2L_S/a}{3(1+2L_S/a)}\left(1 + \kappa a + \frac{(\kappa a)^2}{2}\exp(\kappa a)E_5(\kappa a)\right)\right)} \tag{3}$$

This equation accounts for the fact that the double layer is not sufficiently thin, as the ratio between particle radius, $a$ (measured to be 65 nm, see **Figure 2a** above) and the Debye length, $\kappa^{-1}$, (defined by eq. (4) below), varies between 4.5 ($I = 0.45$ mM) and 67 ($I = 100$ mM). The equation also considers that the studied particles are weakly charged, leading to a non-slip boundary condition on the particle surface. In eq. (3), $\kappa$ is the Debye parameter which is related to the ionic strength, $I$, by the following relation [32]:

$$\kappa = \left(\frac{2N_A e^2 I}{\varepsilon \varepsilon_0 k_B T}\right)^{1/2}, \tag{4}$$

where $N_A$ is the Avogadro number, $e$ is the elementary charge, $I$ is the ionic strength expressed in mM, $k_B$ is the Boltzmann constant, $T$ is the temperature, and $L_S$ is the slip length. $E_n(x)$ is the $n^{\text{th}}$ order exponential integral defined as $E_n(x) = \int_1^\infty \frac{\exp(-tx)}{t^n} dt$.

To determine the value of the slip length $L_S$, we used the data shown in Figure 2 in Ref. [45]. This data shows that $L_S$ depends almost linearly on the distance between neighboring charges. Assuming that the surface charge density is 2 mC/m$^2$, as estimated by eq. (2), we calculated that the distance between neighboring charges on the particle surface is 8.9 nm and the slip length is 4.2 nm. Using eq. (3) with $L_S$ = 4.2 nm and $a$ = 65 nm, we calculated the values of surface potential at ionic strengths at which the determined ζ-potential is above -50 mV. These surface potentials, $\psi_0$, are shown on the right axes of **Figures 2a**, **3c** and **Supplementary Figure S3**. The threshold value of -50 mV for the ζ-potential calculated via Smoluchowksi equation, corresponds to electrophoretic mobility of -4×10$^{-8}$ m$^2$/V.s, which was shown to be the maximal electrophoretic mobility at which eq. (3) gives appropriate description of the experimental data at ionic strengths above 5 mM, see Figure 1 in Ref. [45].

One sees that the estimated values of the surface potentials on the base of eq. (3) are lower in absolute values as compared to the estimated value of the zeta potential on the base of Smoluchowski equation, see e.g. **Figures 2a** and **3c**. This difference is related to the presence of a slip on the boundary of the particles which allow them to move faster as compared to the case when non-slip boundary condition is used (as in the case of Smoluchowski equation).

The calculated values of $\psi_0$ were used to determine the surface charge density, $\sigma_\psi$, of the particles using the following equation [45]:

$$\sigma_\psi = \frac{2k_B T \varepsilon \varepsilon_0 \kappa}{e} \sinh\left(\frac{e\psi_0}{2k_B T}\right) \qquad (5)$$

The measured value of ζ = -20 mV at $I$ = 12 mM NaCl for 2 wt. % S970 corresponds to a surface potential of $\psi_0$ = -8.95 mV according to eq. (3) and to $\sigma_\psi$ = 2.3 mC/m$^2$ according eq. (5). This surface charge density, determined from the ζ-potential measurements, is similar to the value estimated independently from the critical ionic strength for gelation, $\sigma_\psi$ = 2.0 mC/m$^2$, see eq. (2). This good agreement indicates that this approach can be used to determine the surface charge density as a function of pH from measured ζ-potential, shown in **Figure 3c**. The calculated charge densities as a function of pH are shown in **Figure 5b**. These data can be used to determine the adsorption isotherm of hydroxyl ions on the particle surface by using the isotherm described in Ref. [37]:

$$\Gamma = \frac{-\sigma_{\psi}}{e} = \frac{\Gamma_0 v_0 n_0 \exp\left(-\dfrac{\Phi + e\psi_0}{k_B T}\right)}{1 + v_0 n_0 \exp\left(-\dfrac{\Phi + e\psi_0}{k_B T}\right)}, \tag{5}$$

where $\Gamma$ is hydroxyl ions adsorption, $\Gamma_0$ is the saturation adsorption, $v_0 = 1.1 \times 10^{-28}$ m$^3$ is the volume of a hydrated hydroxyl ion in the solution [37], $\Phi$ is the adsorption energy, and $n_0$ is the bulk concentration of the hydroxyl ions, varying with the pH of the solution. In Ref. [37], the adsorption energy of hydroxyl ions was determined to be $\Phi = -25$ $k_B$T from the best fit of the experimental data for hydroxyl ions adsorption on oil-water interface, with $\Gamma_0 = 9.8 \times 10^{-8}$ mol/m$^2$. Our experimental data are well fitted with $\Phi = -28.5$ $k_B$T and $\Gamma_0 = 5.0 \times 10^{-8}$ mol/m$^2$. The very similar results obtained for our particles and for a range of hydrophobic oils in [37] shows that hydroxyl ions are adsorbed on the surface of our particles in a similar manner.

Now we can explain why lowering the pH usually leads to a significant decrease in the critical ionic strength for gelation. For example, the CISG for AlCl$_3$ is determined to be 4.2 mM for 4 wt. % S970, whereas it is 6 mM for NaCl, see **Table 1**. This difference is related to the different pH values of the solutions: pH = 6.5 for NaCl-containing solution and pH = 5.0 for AlCl$_3$-containing solution. Therefore, the lower pH in the presence of AlCl$_3$, leads to a reduced surface charge density, which in turn lowers the AlCl$_3$ concentration required to induce gelation.

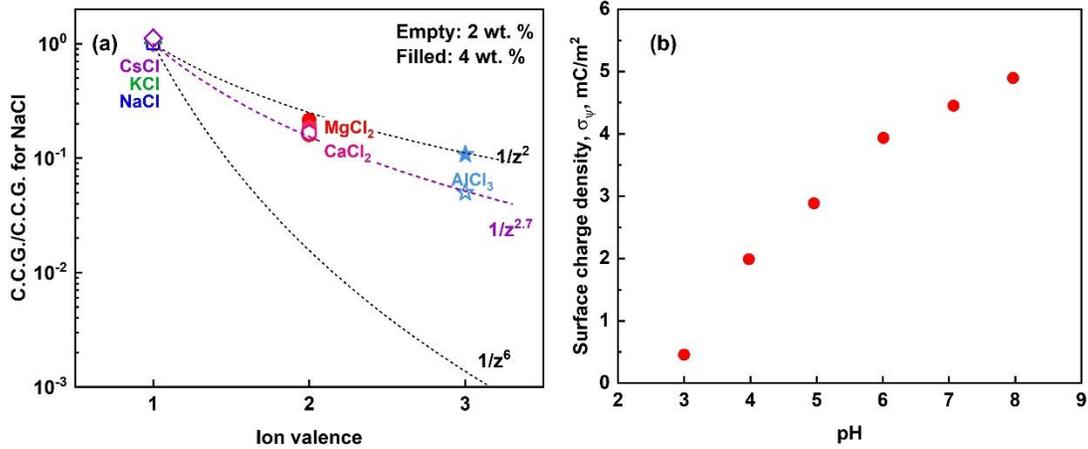

**Figure 5.** (a) Effect of different electrolytes over the phase behavior of S970 dispersions. Critical concentration for gel formation for NaCl, KCl, CsCl, MgCl$_2$, CaCl$_2$ and AlCl$_3$ divided by the critical concentration for NaCl. The data are presented as a function of the ion valence, z. The concentrations follow the $1/z^{2.7}$ dependence for 2 wt. % S970 samples (empty symbols), whereas the exponent decrease slightly for the 4 wt. % samples (filled symbols). $1/z^2$ and $1/z^6$ lines are plotted for comparison. (b) Surface charge density of the studied particles at $I = 5$ mM as a function of pH.

The critical salt concentrations for phase separation beyond which the formation of gels was impossible were also investigated for 2 wt. % S970. They are summarized in **Table 1**. The c.c.p. for the monovalent ions followed the order of $Cs^+ < K^+ < Na^+$. The observed sequence is in good agreement with the molar Gibbs energy of ion hydration, see **Table 1** [42]. This order also coincides with the Hofmeister series [46], suggesting specific interactions between the cations and the hydroxyl anions adsorbed on the diester particles surface.

For the currently studied systems, these "specific interactions" are probably related to the ion affinity to hydration and respectively, how easy would be for a given ion to lose the water molecules present in its hydration shell. In particular, the hydrated radius of the monovalent ions studied is $\approx 0.33 - 0.36$ nm, see **Table 1** [32]. Therefore, at low salt concentrations, when the potential is still high enough, the ions will be probably present within the diffuse electric double layer where the hydration shell will be still preserved. However, once the salt concentration increases, to further decrease the electrostatic potential, the ions will need to enter into the *Stern* layer and loose (most of) the water molecules in their hydration shell. Therefore, this would be the easiest for the poorly hydrated $Cs^+$. In contrast, $Na^+$ are more hydrated [47] and the gels are able to sustain higher concentration of NaCl before they collapse. As shown in Ref. [48], the specific adsorption energy of counterions on the air-water interface or onto the micellar surfaces is related to the ability of counter ions to lose their hydration shell. The values of the radii of the bare and hydrated ions are used in [48] to calculate the specific interaction energy. It was shown that the specific interaction energy of $Na^+$ is $-0.34 k_B T$, whereas it is $-0.97 k_B T$ for $K^+$. This difference aligns well with the lower concentration of $K^+$ required to induce gel collapse compared to $Na^+$. Additionally, the molar Gibbs energy of ion hydration increases from $Ca^{2+}$ to $Mg^{2+}$, which is in good agreement with lower concentration of $Ca^{2+}$ needed to induce precipitation as compared to $Mg^{2+}$. However, a general trend for c.c.p. dependence on the ion valency could not be determined as the experimental points remained relatively scattered, c.c.p. $\propto z^{-2}$ for $MgCl_2$, whereas c.c.p. $\propto z^{-3.4}$ for $CaCl_2$, and c.c.p. $\propto z^{-4.6}$ for $AlCl_3$ at 2 wt. % S970 concentration.

After showing that the qualitative phase behavior of S970 dispersions always follows the same sequence upon increase of the salt concentration: freely flowing low viscosity sample → viscous liquid sample approaching the percolation threshold (perturbed gel) → non-flowing gel → highly viscous, but flowing collapsed gel → phase separated system, next we investigated whether the type of electrolyte used to decrease the electrostatic repulsion between the particles affects the rheological properties of the prepared gels.

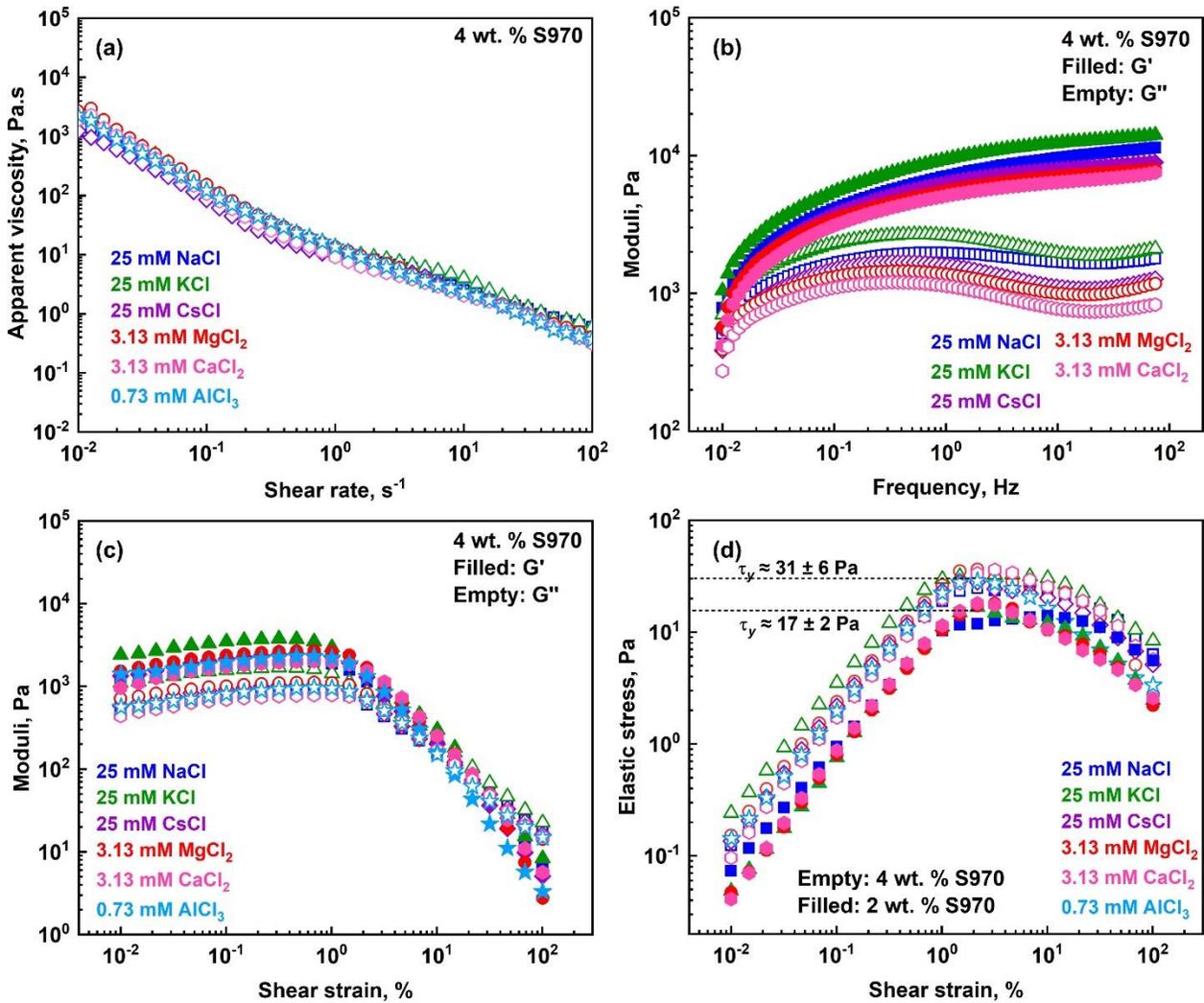

**Figure 6. Rheological properties of S970 gels prepared with different electrolytes.** (a) Apparent viscosity as a function of shear rate for 4 wt. % S970 samples. (b,c) Viscoelastic properties of gels for 4 wt. % S970 samples. Filled symbols: G', empty symbols: G''. (d) Elastic stress as a function of shear rate determined for 2 wt. % (filled symbols) and 4 wt. % (empty symbols) S970 gels. The average yield stresses are denoted on the graph. Different symbols denote the salt which has been used for the preparation of the gels: blue squares – 25 mM NaCl; green triangles – 25 mM KCl; purple rhombus – 25 mM CsCl; red circles – 3.13 mM MgCl$_2$; pink hexagons – 3.13 mM CaCl$_2$; light blue stars – 0.73 mM AlCl$_3$. The experimental data shown in the figures are averaged from three measurements.

Properties of the gels prepared with 25 mM monovalent salts (NaCl, KCl and CsCl), 3.13 mM divalent salts (MgCl$_2$ and CaCl$_2$, $I$ = 9.39 mM), and 0.73 mM AlCl$_3$ ($I$ = 4.38 mM) were compared at 2 wt. % and 4 wt. % S970 surfactant concentrations. The concentrations of the divalent and trivalent salts were chosen to be ca. 10-20% higher than the c.c.g. determined for the specific salt. These concentrations are approximately equal to the NaCl concentration divided by $2^3$ and $3^3$, respectively, as suggested by the results shown in **Figure 5a**.

The obtained results are shown in **Figure 6** and **Supplementary Figure S7**. The viscosities and viscoelastic properties of the gels prepared with different salts were practically the same. Furthermore, the conclusions made for the NaCl containing gels discussed in Section 3.4 above confirmed also for the gels prepared with the other salts. Similar behavior was also observed regarding the maximum shear strain at which the gel flowing began, see **Figure 6d**. This demonstrates that the gel properties are primarily determined by the interactions between the structural units within the gels, i.e. the diester particles and aggregates. These properties are not specific to the electrolyte used, which serves mainly to decrease the electrostatic repulsion caused by the adsorption of hydroxyl anions. Gels with identical characteristics can be obtain with various electrolytes when the counterion valence is accounted, as well as the concentration range for gel formation and the position of the specific electrolyte concentration chosen within it. When the concentration of the salt is increased, it can be expected that the viscosities and yield stresses of the prepared gels will also increase, as shown for NaCl in **Figures 1** and **4**. However, once the electrostatic interactions vanish completely, the gels are no longer stable and phase separation occurs.

### 3.6. Gel formation in surfactant systems and suspensions – discussion

The currently described gel formation differ significantly from the previously known mechanisms by several aspects: (1) aqueous nonionic surfactant solutions are used at relatively low surfactant concentrations, 1.3 wt. % to 4 wt. % (which is equivalent to $\approx$ 17.7 and 54.5 mM, respectively); (2) the nanometer-sized particles causing the formation of space-spanning percolated network are formed in-situ upon dissolution of the surfactant into the water; (3) the interaction between the particles are caused by the adsorption of hydroxyl anions on the their surface, a charging mechanism previously shown for various interfaces [37], but not for in-situ prepared colloidal particles. Note that the in-situ obtained nanometer-sized particles in the presently studied system make it more similar to colloidal systems rather than typical solutions of the low-molecular weight surfactants, which are usually described within the paradigm of surface science.

**Table 2.** Rheological properties for the currently studied S970 gels and other colloidal systems investigated previously. Information about the storage modulus in the linear viscoelastic region and zero-shear viscosity, $\eta_0$, (or viscosity measured at a shear rate of 0.01 s$^{-1}$, $\eta_{0.01}$) is compared.

| Concentration, wt. % | System | Storage modulus, G', Pa |
|---|---|---|
| 4 | silica spheres in hexadecane [51] | $1.5 \times 10^1$ |
| 4.7 | EDAB wormlike micelles [52] | $2.0 \times 10^1$ |
| 2 | microgel + 10 wt. % gold nanoparticles [53] | $2.5 \times 10^1$ |
| 10 | SLES + CAPB = 2:1 + 300 mM KCl [54] | $1.3 \times 10^2$ |
| 3 | Laponite dispersion [15,55] | $0.5\text{-}1.8 \times 10^3$ |
| 3-5 | nanocrystalline cellulose hydrogel + 50 mM NaCl [56] | $1\text{-}2 \times 10^3$ |
| **2-4** | **S970 + 25 mM NaCl** | $\mathbf{1\text{-}2 \times 10^3}$ |
| **4** | **S970 + 0.93 mM AlCl$_3$** | $\mathbf{1 \times 10^4}$ |
| 4.2 vol. % (17.5 wt. %) | alumina particles + 1M NaCl [57] | $2 \times 10^4$ |

| Concentration, wt. % | System | Viscosity, $\eta_0$ or $\eta_{0.01}$, Pa.s |
|---|---|---|
| 5 | Spherical nonionic micelles (Brij 58) [20] | $1.3 \times 10^{-2}$ |
| 4-5 | C$_{14}$SME+CAPB or CTAHNC + CTAB = 1:1; wormlike micelles without added electrolyte [58,59] | $2\text{-}4 \times 10^{-3}$ |
| 4 | nanocrystalline cellulose suspension [56] | $1.0 \times 10^{-1}$ |
| 5 | SLES + CAPB = 1:1 + 30 mM MgCl$_2$ wormlike micelles, peak viscosity [59] | $7.5 \times 10^1$ |
| 2.4 | EHSB + 500 mM NaCl [60] | $2.5 \times 10^2$ |
| 10 | SLES + CAPB = 2:1 + 300 mM KCl [54] | $5.3 \times 10^2$ |
| **4** | **S970 + 0.73-25 mM salt, Fig. 6a** | $\mathbf{2 \times 10^3}$ |
| 2 | Laponite dispersion [55] | $5 \times 10^3$ |

Surfactants notation used in the table: EDAB = erucyl dimethyl amidopropyl betaine; SLES = sodium lauryl ether sulfate; CAPB = cocoamido propyl betaine; C$_{14}$SME = tetradecyl sulfonated methyl ester; CTAHNC = cetyltrimethylammonium 3-hydroxynaphthalene 2-carboxylate; CTAB = cetyltrimethylammonium bromide; EHSB = erucamidopropyl betaine

The low surfactant concentration at which gelation occurs is not surprising from a principle viewpoint. It is well established that colloidal aggregation can result in gelation at extremely low particle volume fractions when buoyancy-matched particles are used to avoid sedimentation effects [49]. This phenomenon can occur at arbitrarily low volume fractions if a fractal network spanning throughout the whole space forms, as its density decreases as the size of the network increases [50]. For instance, gels formed from polystyrene colloidal particles with sizes around 19 nm and volume fractions between 0.01% and 0.5% have been reported, though they are very fragile and even a gentle hand shake is sufficient to destroy them [49].

Furthermore, the rheological properties of S970 gels surpass those of entangled wormlike micelle gels formed by ionic surfactants at higher concentrations (5-10 wt. %) [52,54,58-60]. This difference likely arises from the fundamentally distinct structural mechanisms underlying these systems. Unlike typical surfactant gels containing wormlike micelles, S970 gels involve interacting diester particles via H-bonds, rendering their behavior more similar to suspension than to typical solutions of low molecular weight surfactants. Note that strong short-range hydrogen bonds formed between surfactant molecules containing hydrophilic sugar residues (escin saponins) have been previously reported to lead to formation of adsorption layers with significantly enhanced surface elasticity compared to this of the typical low molecular weight surfactants [61]. Therefore, H-bonds can be expected to play a significant role in the phase behavior of other surfactants containing sugar residues in their hydrophilic heads.

To strengthen the analogy between currently studied SE dispersions and the suspensions with solid particles, we included rheological data for suspension in **Table 2**. For instance, 4 wt. % silica spheres dispersed in hexadecane or 2 wt. % microgel particles in the presence of 10 wt. % gold nanoparticles exhibit storage moduli of approximately $20 \pm 5$ Pa. In comparison, the S970 gels achieve storage moduli exceeding $10^3$ Pa. Similar high values have been previously observed in suspensions containing Laponite particles [15,55] or nanocrystalline cellulose rod-like particles [56], see **Table 2**. Therefore, despite being composed entirely of low molecular weight molecules, the S970 gels exhibit viscoelastic properties comparable to those of suspensions containing interacting and/or anisotropic particles.

## 4. Conclusions

Sucrose ester surfactants are nonionic, biodegradable and biocompatible surfactants, widely utilized in cosmetics, foods and pharmaceutics [16-20]. Despite their increasing popularity, the structural organization of SEs in aqueous dispersions (solutions) remains incompletely understood [16,20]. This study investigates the phase behavior of sucrose ester surfactant (S970), which consists of ca. 50% monoesters and 50% diesters, in the presence of electrolytes – an area that has not been explored so far.

The results demonstrated that as the salt concentration increased in 1.3-4 wt. % aqueous surfactant dispersions, the following transitions are observed: freely flowing low viscosity liquids → viscous liquids near the percolation threshold → non-flowing gels → viscous collapsed gels → phase-separated systems. These transitions were found to be governed by electrostatic interactions between diester particles/aggregates formed in-situ during surfactant dissolution [20]. We demonstrated that the reason for the presence of negative charge on particles formed from nonionic molecules is the adsorption of hydroxyl anions on the surface of the particles rather than the presence of free fatty acids, as previously proposed for SE dispersions [21,22]. Gel formation also occurred by reducing hydroxyl ions number through pH decrease.

Notably, gels form across a broad range of conditions: the minimal surfactant concentration needed for gelation was about 1.3 wt. %; pH can be varied between 4.5 and 11.5; salt concentration can be as small as 0.7 mM $AlCl_3$ ($I = 3.5$ mM), 1.2-1.5 mM for divalent salts ($I = 3$-3.75 mM), or 6-9 mM for monovalent salts ($I = 6$-9 mM). The critical gelation concentration depends of ion valence, scaling as $\propto z^{-2}$ to $z^{-3}$, in good agreement with Debye-Hückel theory for poorly charged surfaces, where c.c.c. $\propto z^{-2}$ [41]. We demonstrated that the rheological properties of the gels are independent of the specific electrolyte used. Furthermore, gels quickly restored their initial properties after shearing.

The rheological comparison, summarized in **Table 2**, highlights the significantly higher elastic moduli of S970 gels compared to wormlike micelle-based gels, even at higher concentrations (G' $\approx 10^3$ Pa for 2 wt. % S970 + 25 mM NaCl, whereas G' $\approx 1.3 \times 10^2$ Pa for 10 wt. % SLES + CAPB + 300 mM KCl system containing wormlike micelles [54]). Currently obtained viscosities and moduli are comparable to those for gels formed by attractive colloidal particles, such as nanocrystalline cellulose or Laponite [15,55,56], although only low molecular weight surfactant is used and the interacting particles are formed in-situ upon surfactant dissolution.

The gel-forming mechanism identified in this study introduces new opportunities for the application of green sucrose ester surfactants in a variety of products requiring enhanced elasticity, such as home and personal care products, foods, pharmaceuticals, inks, coatings and more. Moreover, it presents a fundamental advancement in the physical chemistry characterization of the SE aqueous dispersions. A limitation of this study is that not all surfactants can form in-situ nanoparticles. However, similar behavior is expected for surfactants containing a mix of alkyl diesters and monoesters from molecules containing at least several OH-groups that can be esterified, such as glucose, fructose, sorbitol, and polyglycerols, particularly those with long hydrocarbon chains ($C_{16}/C_{18}$). Future research into the effect of monoesters-to-diesters ratios for sucrose esters, effect of the hydrophilic headgroup type, temperature and the presence of other chemical species will be needed to fully understand the potential applications of these novel colloidal gels.

## Acknowledgements

This study was funded by the European Union-NextGenerationEU, through the National Recovery and Resilience Plan of the Republic of Bulgaria, project No BG-RRP-2.004-0008. The support of the Centre of Competence "Sustainable Utilization of Bio-resources and Waste of Medicinal and Aromatic Plants for Innovative Bioactive Products" (BIORESOURCES BG), project BG16RFPR002-1.014-0001, funded by the Program "Research, Innovation and Digitization for Smart Transformation" 2021-2027, co-funded by the EU, is greatly acknowledged. The authors are grateful to Mitsubishi Chemical Corporation for the generous donation of S970 Ryoto$^{TM}$, allowing us to perform this study. The authors are grateful to Mrs. K. Rusanova and Assoc. Prof. L. Mihaylov (Sofia University) for cryo-TEM imaging, and to Mr. V. Georgiev for the laser diffraction measurements.

## CRediT author statement:

D. C. – conceptualization; methodology; investigation; validation; formal analysis; visualization; writing – original draft, review and editing.

N. P. – methodology; investigation.

M. H. – investigation; validation.

S. T. – conceptualization; methodology; supervision; formal analysis; writing – review and editing; funding acquisition.

## Declaration of competing interest

The authors declare that they have no known competing financial interests or personal relationships that could have appeared to influence the work reported in this paper.

## Data availability

Data will be made available on request.

# **Supplementary materials**

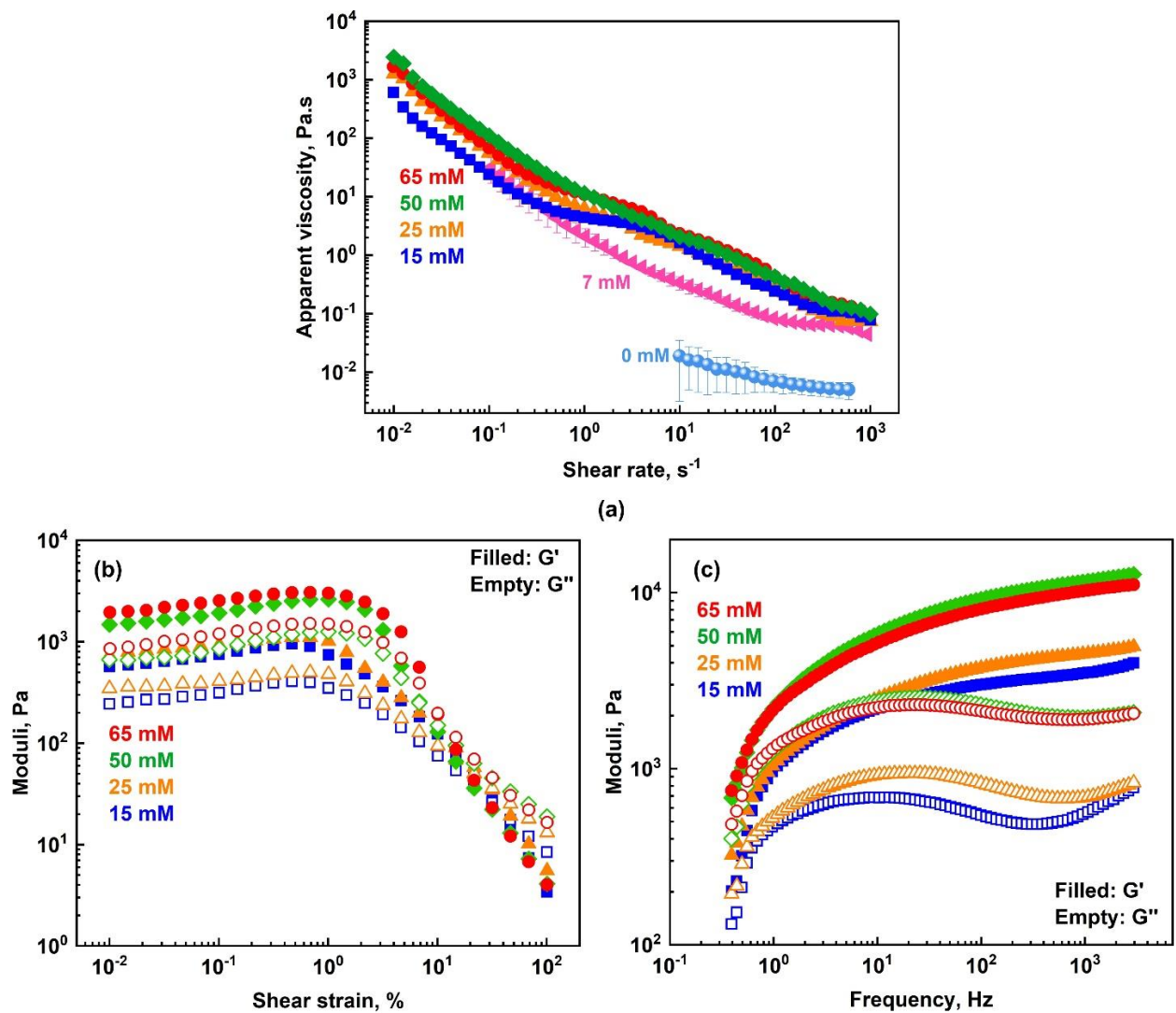

**Supplementary Figure S1. Rheological properties of 2 wt. % S970 samples with different NaCl concentrations.** (a) Apparent viscosity as a function of the shear rate. (b-c) Storage (G', filled symbols) and loss (G'', empty symbols) moduli measured as a function of the applied: (b) shear strain at a fixed frequency of 0.16 Hz; (c) oscillation frequency at a fixed shear strain of 0.5%. Different colors and symbols denote different NaCl concentrations: 0 mM (light blue circles), 7 mM (magenta triangles), 15 mM (blue squares), 25 mM (orange triangles), 50 mM (green rhombus) and 65 mM (red circles). The experimental data shown in the figures are averaged from two independently measured samples.

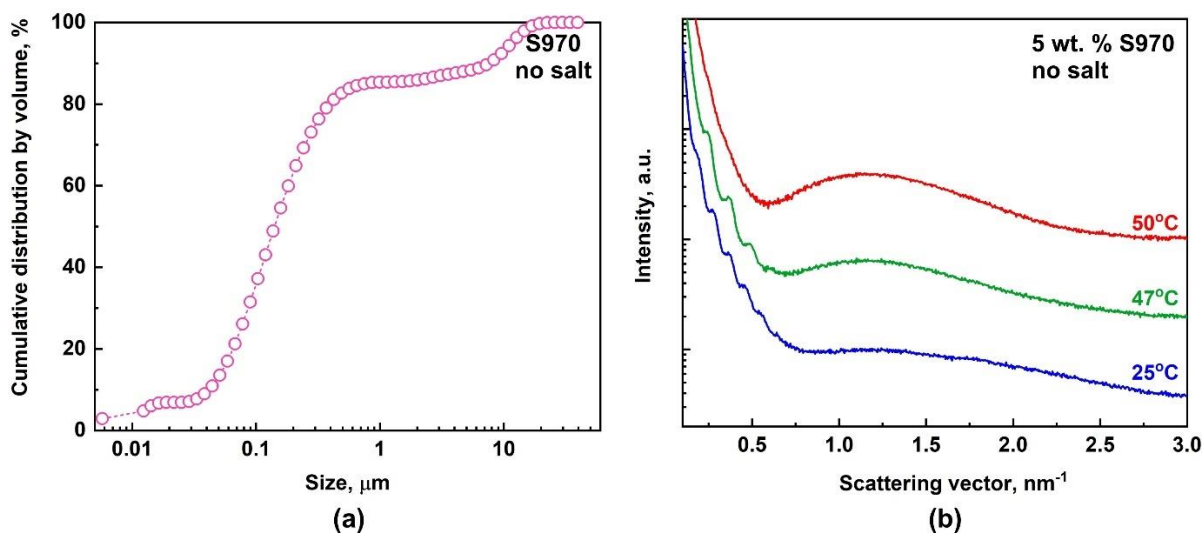

**(a)**  **(b)**

**Supplementary Figure S2.** (a) Cumulative distribution by volume for S970 solutions. Micelles, particles and aggregates are present. (b) SAXS spectra of 5 wt. % S970 solutions without added electrolyte. The spectra have been taken at different temperatures as denoted on the figure. Particles coexisting with short elongated micelles are detected at 25°C. Temperature increase above 47°C leads to disappearance of the particles (shown by the series of peaks at $q < 0.5$ nm$^{-1}$), change of the shape of micelles and increase of their number. The experimental data shown in (A) are averaged from three measurements, whereas the SAXS spectra in (B) are from single measurements performed at different temperatures.

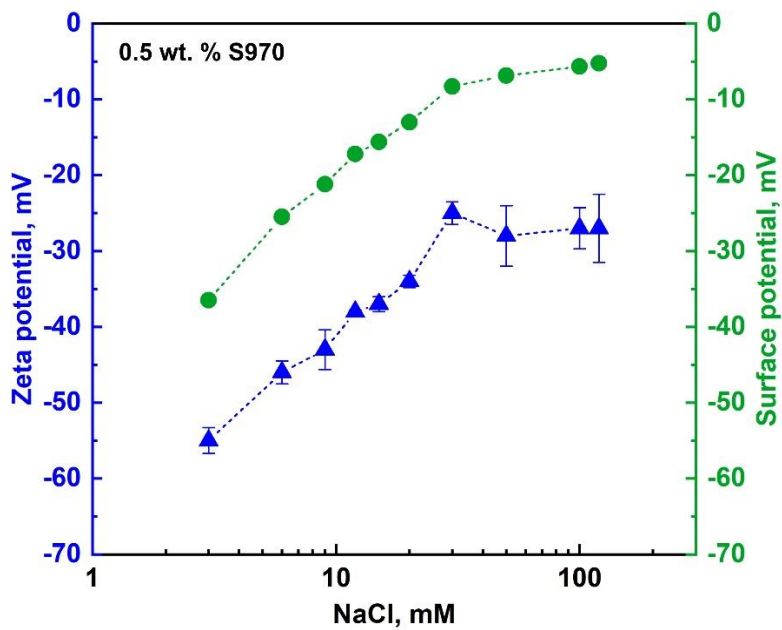

**Supplementary Figure S3.** Zeta potential (blue triangles, left axis) and calculated surface potential via eq. (3) from the main text (green circles, right axis) as a function of the NaCl concentration measured for 0.5 wt.% solutions of S970. The experimental data shown in the figures are averaged from three measurements.

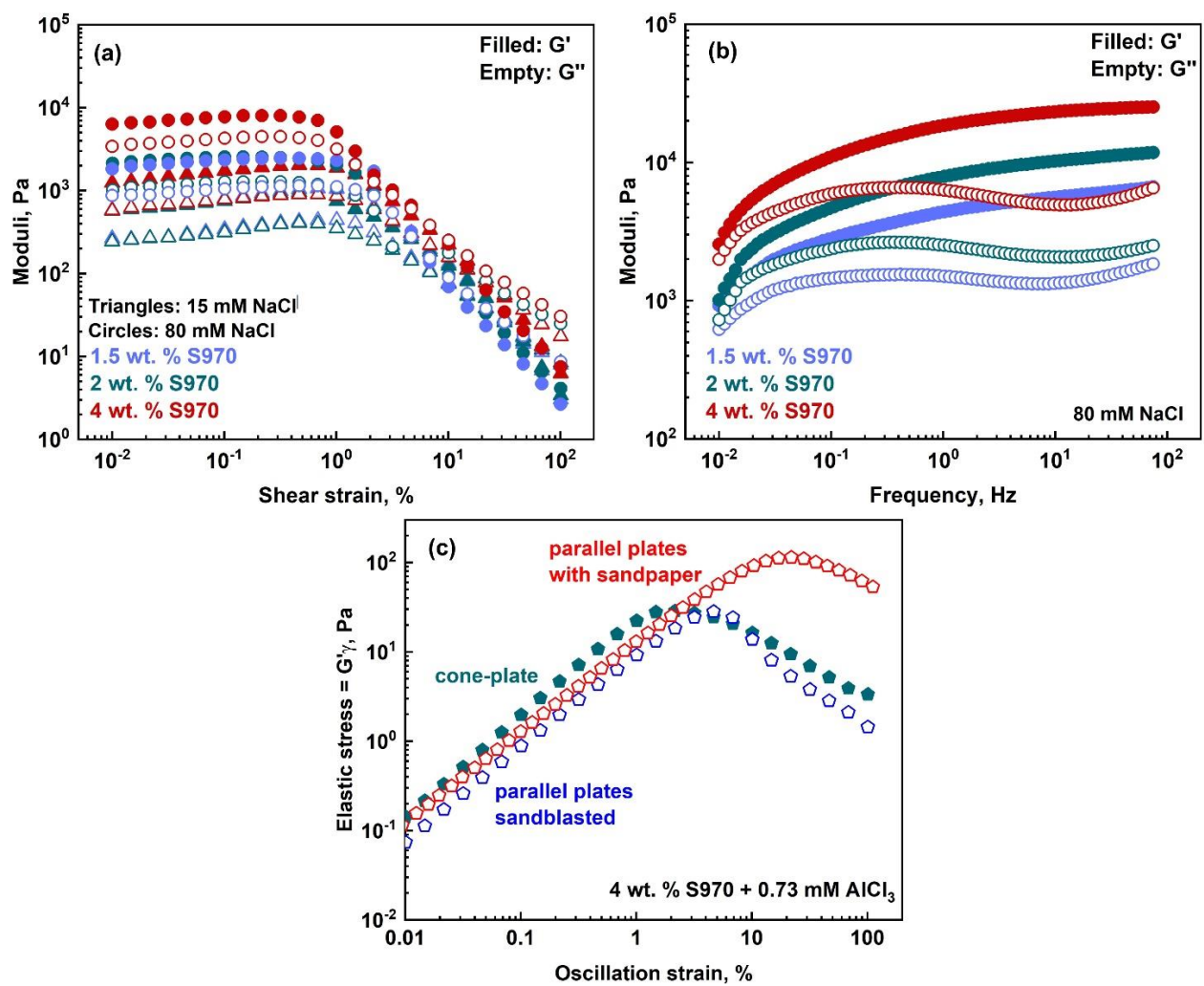

**Supplementary Figure S4. Rheology properties dependence on salt and surfactant concentrations.** (a,b) Storage (filled symbols) and loss (empty symbols) moduli measured as a function of the: (a) applied shear strain at a fixed frequency of 0.16 Hz; (b) applied shear frequency at a fixed shear strain of 0.5%. Different colors denote the surfactant concentrations: 1.5 wt. % - light blue; 2 wt. % - dark green; 4 wt. % - dark red. Data for dispersions prepared with 15 mM NaCl are shown with triangles, whereas the results for samples with 80 mM NaCl are shown with circles. (c) Comparison of the elastic stress determined from the amplitude sweep oscillation measurements for the same sample (4 wt. % S970 + 0.73 mM AlCl$_3$) measured with different experimental set ups: filled dark green symbols – cone and plate geometry; empty blue symbols – parallel plates with sandblasted surface; empty red symbols – parallel plates with attached sandpaper. The initial elastic stress values are within the frame of our experimental accuracy. However, the maximal elastic stress and the oscillation strain at which it is achieved depends strongly on the specific experimental set up due to the wall slip effect. The experimental data shown in the figures are averaged from two measurements.

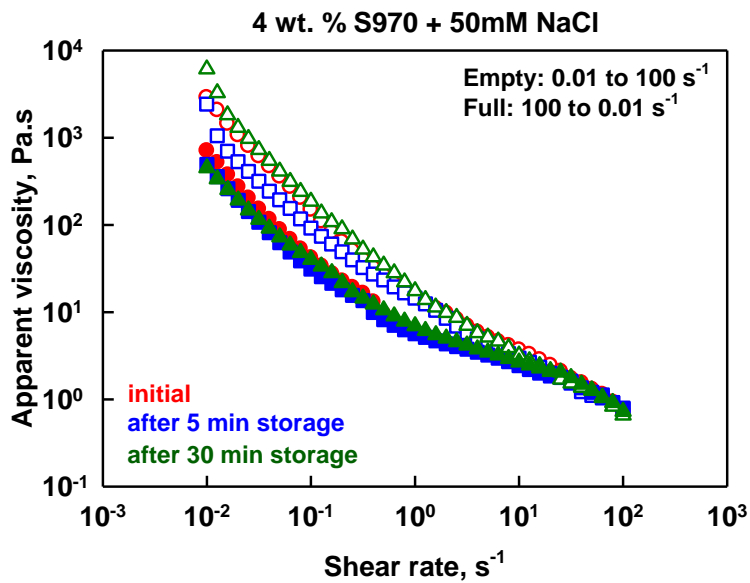

**Supplementary Figure S5.** Apparent viscosity as a function of shear rate for 4 wt. % + 50 mM NaCl: red points represent the experimental results for initial sample; blue points represent the experimental data for the same sample after 5 min of isothermal storage after the first measurement has been performed, and green points represent the results for the same sample after 30 min of storage (after the end of the second measurement). Empty points show the results obtained when the shear rate increases from 0.01 to 100 s$^{-1}$, whereas the full points represent the results obtained when the shear rate decreases from 100 s$^{-1}$ down to 0.01 s$^{-1}$.

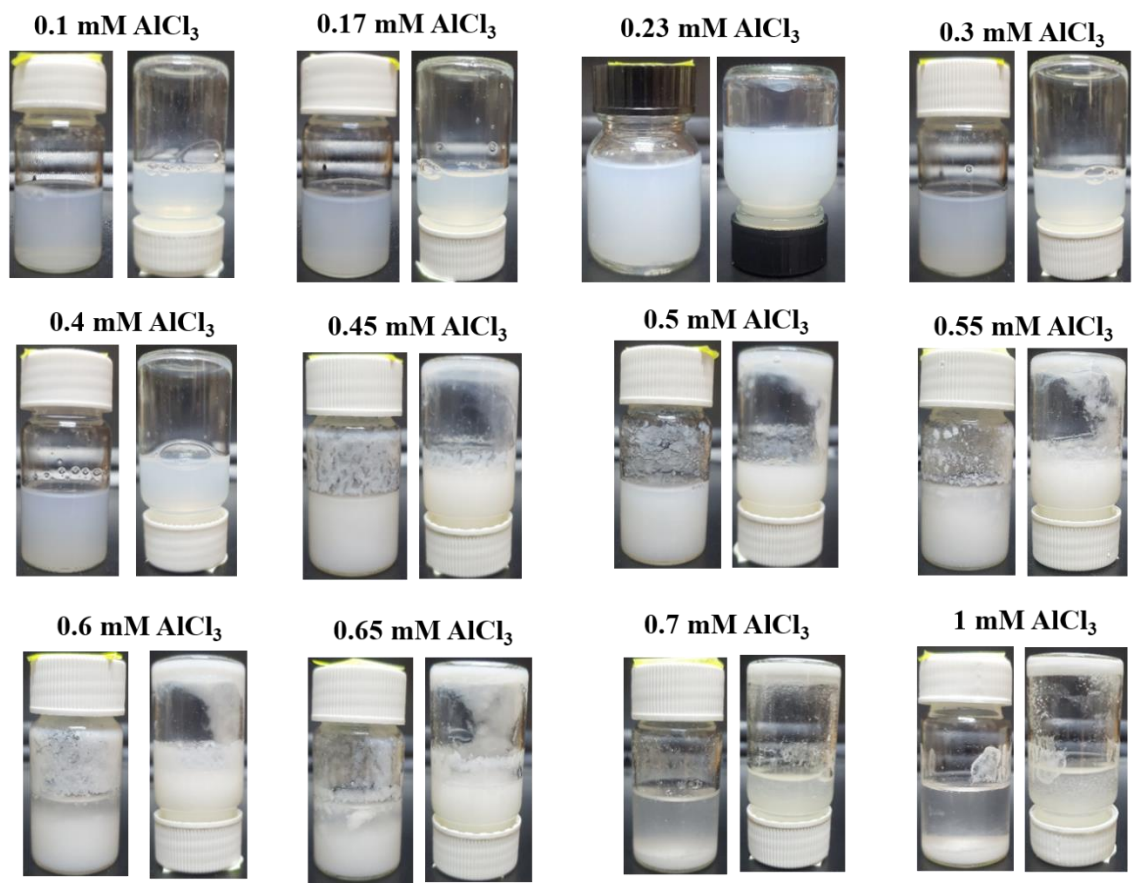

**Supplementary Figure S6.** **Phase behavior dependence of 2 wt. % S970 solutions with added AlCl₃.** Low viscous samples with increasing turbidity are observed for c(AlCl₃) < 0.4 mM. The viscosity increases when the AlCl₃ concentration reach 0.45 mM, however completely non-flowing sample is not obtained. Complete surfactant precipitation and phase separation occurs when c(AlCl₃) ≥ 0.7 mM.

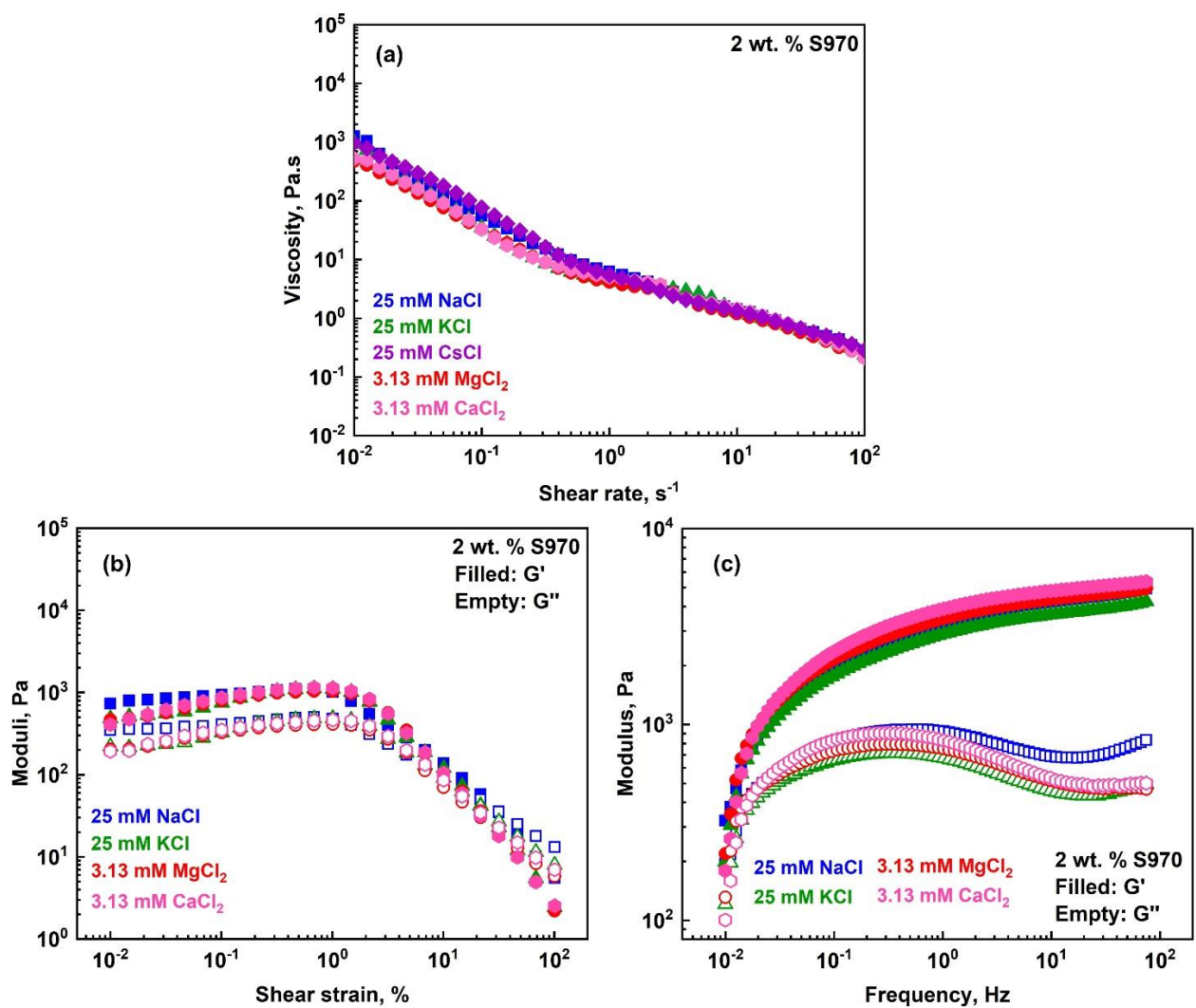

**Supplementary Figure S7. Rheological properties of 2 wt. % S970 gels prepared with different salts.** (a) Apparent viscosity as a function of the shear rate. (b-c) Storage (filled symbols) and loss (empty symbols) moduli measured as a function of the: (b) applied shear strain, (c) frequency. Different colors denote different salts: blue – 25 mM NaCl; green – 25 mM KCl; purple – 25 mM CsCl; red – 3.13 mM $MgCl_2$; pink – 3.13 mM $CaCl_2$. The experimental data shown in the figures are averaged from three measurements.